\documentclass[sigconf, screen]{acmart}

\usepackage[utf8]{inputenc}

\copyrightyear{2024}
\acmYear{2024}
\setcopyright{acmlicensed}\acmConference[ASE '24]{39th IEEE/ACM International Conference on Automated Software Engineering }{October 27-November 1, 2024}{Sacramento, CA, USA}
\acmBooktitle{39th IEEE/ACM International Conference on Automated Software Engineering (ASE '24), October 27-November 1, 2024, Sacramento, CA, USA}
\acmDOI{10.1145/3691620.3695472}
\acmISBN{979-8-4007-1248-7/24/10}

\usepackage{tabu}
\usepackage{graphicx}
\usepackage{textcomp}
\usepackage{xcolor}
\usepackage[ruled,linesnumbered]{algorithm2e}
\usepackage{algpseudocode} 
\usepackage{subcaption}
\usepackage{url}
\usepackage{multirow}
\usepackage{framed}
\usepackage{adjustbox}
\usepackage{hyperref}
\usepackage{bbm}
\usepackage{xspace}
\usepackage{tcolorbox}
\tcbuselibrary{most}

\usepackage{amsmath,amsfonts,bm}

\def\eqref#1{equation~(\ref{#1})}

\def\1{\bm{1}}

\def\vx{{\bm{x}}}

\DeclareMathAlphabet{\mathsfit}{\encodingdefault}{\sfdefault}{m}{sl}
\SetMathAlphabet{\mathsfit}{bold}{\encodingdefault}{\sfdefault}{bx}{n}

\DeclareMathOperator*{\argmax}{arg\,max}

\usepackage{marvosym}
\tcbuselibrary{most}

\usepackage{bbding} 
\usepackage{textcomp}
\usepackage{booktabs}
\usepackage{multirow}
\usepackage{bbm}

\usepackage{caption}

\usepackage{xspace}

\newcommand{\tool}{\textsc{FAST}\xspace}

\newcommand{\commentout}[1]{}

\def\vx{{\bm{x}}}

\hypersetup{hidelinks,
	colorlinks=true,
	allcolors=black,
	pdfstartview=Fit,
	breaklinks=true}

\usepackage{tcolorbox}
\usepackage{verbatim}
\usepackage{marginnote}

\usepackage{soul}

\title{\tool: Boosting Uncertainty-based Test Prioritization Methods for Neural Networks via Feature Selection}

\author{Jialuo Chen}
\authornote{Work done while at the University of Oxford.}
\affiliation{%
  \institution{Zhejiang University}
    \city{Hangzhou}
  \country{China}
  }
\email{chenjialuo@zju.edu.cn}

\author{Jingyi Wang}
\authornote{Co-corresponding authors.}
\affiliation{%
  \institution{Zhejiang University}
  \city{Hangzhou}
  \country{China}
  }
\email{wangjyee@zju.edu.cn}

\author{Xiyue Zhang}
\affiliation{%
  \institution{University of Oxford}
  \city{Oxford}
  \country{United Kingdom}
  }
\email{xiyue.zhang@cs.ox.ac.uk}

\author{Youcheng Sun}
\affiliation{%
  \institution{University of Manchester}
  \city{Manchester}
  \country{United Kingdom}
  }
\email{youcheng.sun@manchester.ac.uk}

\author{Marta Kwiatkowska}
\affiliation{%
  \institution{University of Oxford}
  \city{Oxford}
  \country{United Kingdom}
  }
\email{marta.kwiatkowska@cs.ox.ac.uk}

\author{Jiming Chen}
\affiliation{%
  \institution{Zhejiang University}
  \city{Hangzhou}
  \country{China}
  }
\email{cjm@zju.edu.cn}

\author{Peng Cheng}
\authornotemark[2]
\affiliation{%
  \institution{Zhejiang University}
  \city{Hangzhou}
  \country{China}
  }
\email{lunarheart@zju.edu.cn}

\begin{document}

\begin{abstract}

   Due to the vast testing space, the increasing demand for effective and efficient testing of deep neural networks (DNNs) has led to the development of various DNN test case prioritization techniques. However, the fact that DNNs can deliver high-confidence predictions for incorrectly predicted examples, known as the over-confidence problem, causes these methods to fail to reveal high-confidence errors.
   To address this limitation, in this work, we propose \tool, a method that boosts existing prioritization methods through guided \textbf{F}e\textbf{A}ture \textbf{S}elec\textbf{T}ion. \tool is based on the insight that certain features may introduce noise that affects the model’s output confidence, thereby contributing to high-confidence errors. It quantifies the importance of each feature for the model's correct predictions, and then dynamically prunes the information from the noisy features during inference to derive a new probability vector for the uncertainty estimation. With the help of \tool, the high-confidence errors and correctly classified examples become more distinguishable, resulting in higher APFD (Average Percentage of Fault Detection) values for test prioritization, and higher generalization ability for model enhancement. We conduct extensive experiments to evaluate \tool across a diverse set of model structures on multiple benchmark datasets to validate the effectiveness, efficiency, and scalability of \tool compared to the state-of-the-art 
   prioritization techniques.

\end{abstract}



\maketitle

\section{Introduction}

Deep neural networks (DNNs) have demonstrated remarkable performance in diverse safety-critical domains such as self-driving cars~\cite{deeptest} and malware detection \cite{yuan2014droid}. Such domains have a low tolerance for mistakes. Compared to traditional code-driven software, DL-based systems adopt a data-driven programming paradigm, and thus introduce 
reliability challenges related to data faults (e.g., mispredicted data samples). Therefore, it is essential to conduct a high-quality test for DNNs to expose the hidden vulnerabilities as much as possible.
However, testing for DNNs generally requires massive labeled data, while building test oracles by manually labeling for a large test set can be costly~\cite{nns}, especially for tasks that require domain-specific knowledge (e.g., medical images).

To increase the efficiency of DNN testing and reduce the labeling cost, it is essential to prioritize the test cases and identify a set of high-quality examples to reveal the prediction faults. A variety of test case prioritization methods~\cite{deepgini,prima,shen2020multiple,xie2022boosting,certpri} have been proposed based on carefully designed testing metrics, such as neuron coverage (NC)~\cite{deepxplore,deepgauge,concolic}, surprise adequacy (SA)~\cite{surprise,weiss2021review}, and prediction uncertainty~\cite{deepgini,byun2019input,maxp}. Recent studies~\cite{weiss2022simple,nns,maxp} have validated that the prioritization methods based on uncertainty achieved dominant results and less overhead. 
These methods (e.g., DeepGini~\cite{deepgini}) calculate the uncertainty score directly from the model's probability output, with the assumption that test cases with higher uncertainty scores (closer to the decision boundary) are more likely to be prediction errors. However, DNNs can be very confident on the wrongly predicted examples~\cite{nguyen2015deep,guo2017calibration} (the over-confidence problem), which causes these errors to be naturally overlooked by existing uncertainty-based methods. 
In practical applications, especially those heavily dependent on high-confidence decisions, over-confident errors can be more troublesome. Recently, nearest neighbors smoothing (NNS)~\cite{nns} has been proposed to help mitigate this problem by averaging the outputs with a set of closely matched samples, showing notable improvements. However, its performance largely depends on the quality of the neighbors and can be computationally costly due to the search algorithm.

Since DNNs make the final prediction of an input based on the extracted features (the outputs of neurons), there should be connections between certain characteristics of hidden features and the prediction results, whether correct or erroneous. 
Despite considerable efforts within the software engineering community to characterize the behaviors of neurons, such as neuron activation coverage metrics~\cite{deepxplore,deepgauge,concolic}, several studies ~\cite{deepgini,limit,yan2020correlations,wang2021robot} have found that these measurements are not necessarily even negatively indicative of bug-revealing capabilities. We suppose that existing metrics focus solely on analyzing output values of intermediate layers; however, due to the high non-linearity of DNNs, their actual contribution to final predictions has not yet been clearly examined.
Intuitively, different features are combined together to describe specific groups of examples (e.g., `cat' or `dog'), each contributing differently to the class prediction. 
By examining the intermediate features of the model and empirically measuring each feature's contribution, we show that there is a certain portion of `redundant' features having minimal impact on the correctly predicted examples. On the contrary, such features can undesirably introduce noisy information aggregated to the model’s final output, thereby leading to high-confidence errors.

\begin{figure}[t]
\centering
\includegraphics[width=0.4\textwidth]{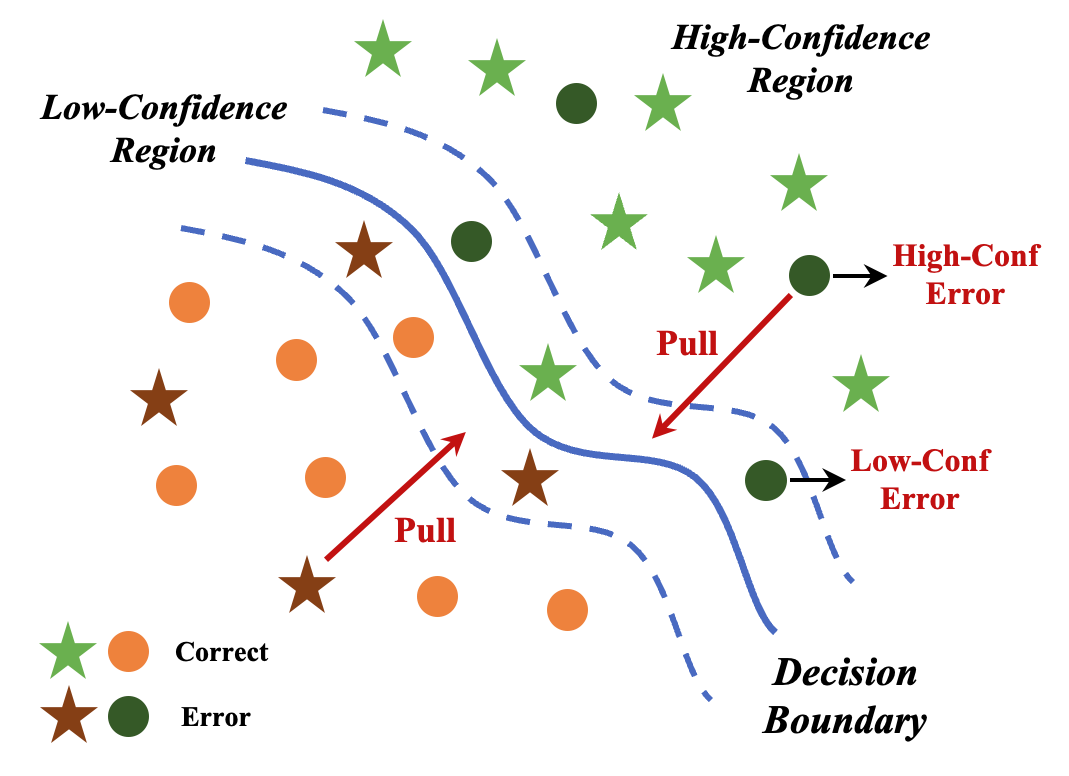}
\vspace*{-2mm}
\caption{Illustration of the key limitation of existing uncertainty-based methods, which focus primarily on the test cases close to the decision boundary while neglecting the high-confidence errors. \tool helps to expose such high-confidence errors by dynamically suppressing their confidence, pulling them towards the low-confidence region.}
\label{fig:curve}
\vspace*{-1mm}
\end{figure}

To effectively reveal such high-confidence errors, we propose a simple yet effective method for test input prioritization, namely FAST. It prioritizes test inputs by integrating guided feature selection during inference, so as to remove the potential noise incorporated in the output mispredictions for uncertainty measure. Specifically, \tool selectively prunes a subset of `noisy' features from a given feature layer based on their contribution measurements, then uses the remaining features to derive the probability vector for uncertainty estimation. As a result, the original high-confidence errors will tend to be given a lower confidence score, making them easier to prioritize based on the calibrated probability vector, as illustrated in Fig.~\ref{fig:curve}. \tool positions itself as a plug-and-play technique that can be easily applied with existing uncertainty-based test prioritization methods. We extensively evaluated \tool on 7 benchmark datasets (including image, text, and audio data) with 9 different DNN structures (including both CNN and RNN structures) 
against 10 prioritization baselines from different types. The results confirm the effectiveness, efficiency, and scalability of \tool compared to the state-of-the-art methods. In terms of the overall prioritization performance, \tool achieves 3.19\% higher APFD (Average Percentage of Fault Detection) values than uncertainty-based methods and 1.78\% higher than recent work (NNS~\cite{nns})  on average. Specifically, for small selection budgets, \tool exposes over 13.63\% more errors than uncertainty-based methods and 8.16\% more than NNS, respectively. 
In terms of model enhancement guided by test prioritization, \tool is effective in enhancing the model performance by rectifying more prediction errors through retraining (with 5\% selection budget), achieving an accuracy improvement of 3.47\% on average, which is 13.36\% higher than NNS.
Notably, \tool only incurs a slight additional time cost compared to the uncertainty-based methods, making it more efficient than most existing methods.

In summary, we make the following contributions.
\begin{itemize}

\item We propose and implement a novel prioritization method \tool through feature selection (noisy feature pruning). With \tool, more high-confidence errors can be exposed compared to traditional uncertainty-based methods.

\item We extensively evaluate \tool on multiple common benchmark datasets with various model structures, validating the superior performance in prioritizing test cases compared to previous work. As a lightweight and generic method, \tool also shows high efficiency and scalability.

\item We release \tool at \cite{code} to facilitate future studies in this area.

\end{itemize}

\begin{figure*}[]
\centering
\includegraphics[width=0.85\textwidth]{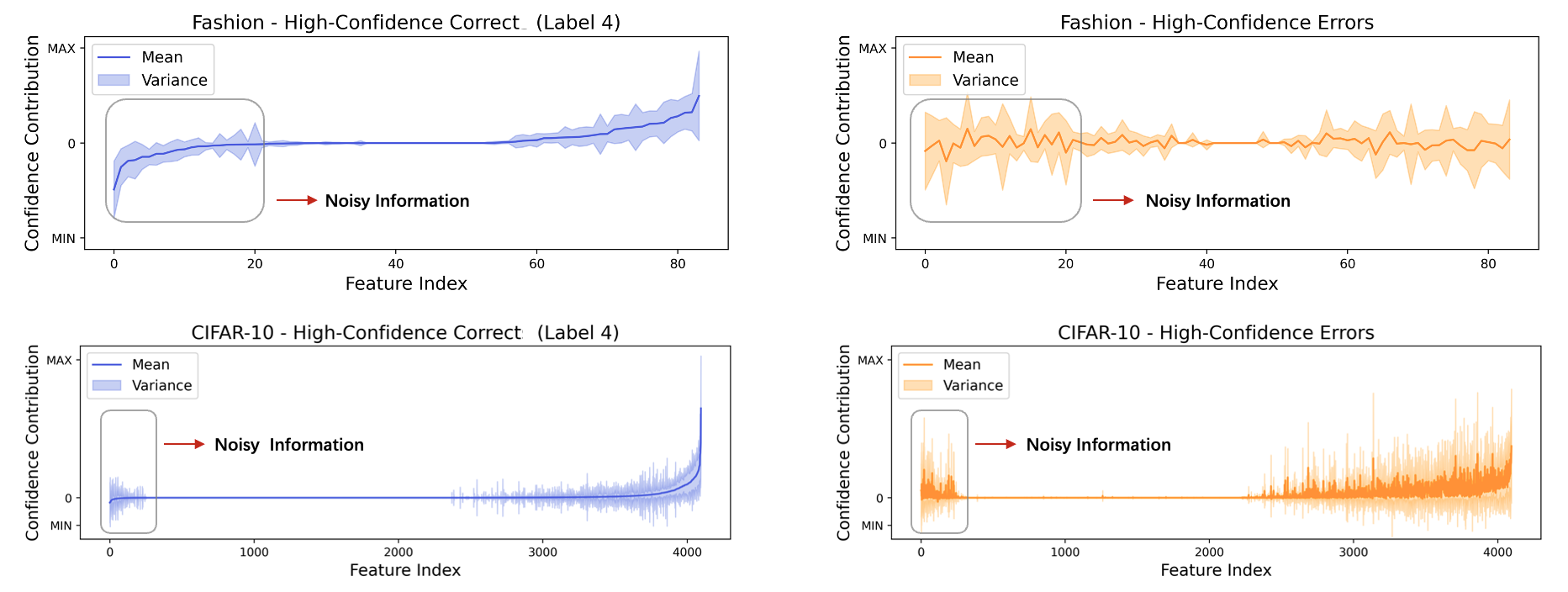}
\vspace*{-2mm}
\caption{The feature contribution to the output confidence for the FASHION model (LeNet-5) and the CIFAR-10 model (VGG-16). The features are sorted in an increasing order based on their average contribution over the correctly classified examples with high confidence. The key facts are: 1) the most important features at the higher end significantly impact both correctly and incorrectly classified samples with high confidence, and 2) while the features at the lower end contribute the least to high-confidence correct classifications, they can still significantly affect high-confidence misclassifications.}
\label{fig:featurecontribution}
\end{figure*}

\section{Background}
\label{sec:background}

\subsection{Deep Neural Networks}
\label{bg:dnn}
We focus on deep neural networks (DNNs) for classification. Typically, a DNN classifier~\cite{goodfellow2016deep} is a decision function $f: X\rightarrow Y$ mapping an input $\vx \in X$ to a label $y \in Y=\{1, 2, \cdots, C\}$, where $C$ is the number of classes. It comprises of $L$ layers: $\{f^1, f^2, \cdots, f^{L-1}, f^L\},$ where $f^1$ is the input layer, $f^L$ is the probability output layer, and $\{f^2,\cdots,f^{L-1}\}$ are the hidden layers. Each layer $f^{l}$ can be denoted by a collection of neurons: $\{n_{l,1}, n_{l,2}, \cdots, n_{l,N_l}\},$ where $N_l$ is the total number of neurons at that layer. Each neuron is a computing unit that computes its output by applying a linear transformation followed by a nonlinear operation to its input (i.e., output from the precedent layer). We use $n_{l,i}(x)$ to denote the function
that returns the output of neuron $n_{l,i}$ for a given input $ x \in X$. Then, we have the output vector of layer $f^{l} $ ($2 \leq l \leq L$): $f^{l}(x) = \left\langle  n_{l,1}(x), n_{l,2}(x), \cdots, n_{l,N_l}(x) \right\rangle$. Specifically, we use $p(x)$ to denote the output probability vector of the final layer, then, the predicted label $f(x)$ is computed as $f(x)=\argmax p(x)$.

\subsection{Test Case Prioritization}
To improve the DNN testing efficiency and reduce the labeling cost, a variety of test case prioritization methods have been proposed.  
For a given model $f$ to test and a test suite $T$, the purpose of test case prioritization is to prioritize the test cases in $T$ in a guided way, and when the test stops at a certain point, it is expected that the executed test cases (often selected according to the rank) will expose as many faults as possible. For a classification model, a test case $x\in T$ is identified as a fault when the model's predicted label $f(x)$ is inconsistent with its ground-truth label.  
The core of test case prioritization is to quantify the possibility of a test case being a fault-trigger. Based on the required information, existing prioritization methods can be classified into two classes: 1) internal-based and 2) prediction-based (or uncertainty-based). 

\vspace{-1mm}
\subsubsection*{Internal-based Prioritization} 
As the model prediction is based on the propagation of layer outputs, it is natural to analyze the model internal status (i.e., hidden feature outputs) to provide evidence for judging a potential error. Multiple coverage metrics such as Neuron Activation Coverage (NAC)~\cite{deepxplore} and Neuron Boundary Coverage (NBC)~\cite{deepgauge} have been proposed to depict the covered neuron status for guiding the testing procedure. A test case that achieves a higher coverage is regarded as a more valuable input and will be prioritized. Another group of metrics utilizing the model internal is Surprise Adequacy (SA)~\cite{surprise}, which measures how surprising an input is, i.e., how far is a given test case from the training set. Specifically, Likelihood-based Surprise Adequacy (LSA) utilizes the kernel density estimation (KDE) to obtain a density function from the training instances, which allows the direct estimation of the likelihood of a test case being a fault trigger. Distance-based Surprise Adequacy (DSA) measures the distance between the test case and the closest example in the training dataset. 
However, several studies have found that all the above coverage-based criteria are not necessarily or even negatively indicative of fault-revealing 
capabilities~\cite{deepgini,limit,yan2020correlations,quote}.

\vspace{-1mm}
\subsubsection*{Uncertainty-based Prioritization} 
Instead of utilizing the internal outputs, uncertainty-based methods solely rely on the model's final outputs, i.e., the probability vector $p(x)$. The intuition behind is that, if a DNN model predicts a test instance with nearly equal probability values on all candidate classes, indicating the model is uncertain (or confused) on predicting this test case, then it is more likely to be an incorrect prediction. We briefly introduce representative uncertainty metrics that are used in uncertainty-based test prioritization methods.
DeepGini~\cite{deepgini} quantifies the likelihood of a test case $x$ being incorrectly predicted as: $Gini(x)= 1-\sum_{i=1}^{C}p_i^2(x)$, where $p_i(x)$ is the prediction probability of class $i$. A high Gini index impurity represents high uncertainty.
MaxP~\cite{maxp} directly uses the highest probability output as: $MaxP(x)=\max p_i(x)$ as the metric. Margin~\cite{scheffer2001active} is another way for quantifying prediction uncertainty, which is the difference between the highest and the second highest prediction probabilities, and a low margin represents a high uncertainty. 
After quantifying the uncertainty, the test cases with high uncertainty (close to the decision boundary) will be prioritized. 
Compared to internal-based methods, uncertainty-based methods are more lightweight and achieve better results~\cite{weiss2022simple,hu2021towards}, as they do not require recording a heavy profile of neuron activations.

\vspace{-1mm}
\subsubsection*{Over-confidence Problem of DNNs} 
Recent studies~\cite{deepgini,weiss2022simple,nns} have demonstrated the dominance of simple uncertainty-based methods in prioritizing the faults of DNNs, However, they can be further improved. Modern DNN models could suffer from the over-confidence problem, in that they tend to give high probability scores to the wrong predictions, due to model capacity, insufficient training data, or other factors. In such cases, the uncertainty-based methods will fall short of exposing and prioritizing these high-confidence faults. As illustrated in Fig.~\ref{fig:curve}, uncertainty-based methods will only prioritize the test cases close to the decision boundary (in the dotted low-confidence region), while neglecting the remote errors that are far away from the decision boundary. In this work, we propose \tool, a method to improve the uncertainty-based prioritization methods to uncover more high-confidence errors.

\section{Methodology}
In this section, we first present the motivation behind and then give details of each part of our prioritization method \tool.

\subsection{Motivation}
Considering the fact that the final prediction results are made based on a group of extracted features, we investigate the over-confidence problem faced by existing prioritization methods from the perspective of features. For a given class, different features (representing different learned patterns) have different levels of `contribution' to the predictions. 
Instead of utilizing the numerical output values of the neurons to quantify the contribution (e.g., activation value~\cite{deepxplore} and activation frequency~\cite{tian2020testing}), we measure it in a more intuitive way, that is, the observable difference on the model's final output with and without that feature (we give details in Section~\ref{sub:measurement}). In other words, the feature whose removal causes a significant drop in confidence will be regarded as contributing more, and otherwise as having minimal influence on the final prediction. Then, for a selected feature layer of the model, we calculate the feature-wise contribution for a set of high-confidence correctly and wrongly predicted examples separately. We use correct predictions and errors in the following for simplicity.

As in Fig.~\ref{fig:featurecontribution}, 
there are clear gaps between the feature contributions for supporting the model's correct predictions (indicated by the blue color), with the lowest-ranked features making the least contributions. However, these same low-contributing features could potentially yield a higher contribution towards misleading the predictions  (indicated by the orange color), as they undesirably bring noisy information that is aggregated to the model’s final output, further causing high-confidence errors. These features also show higher variance. Note that the magnitude of contribution can be influenced by the DNN structure, e.g., the VGG network shows higher sparsity than the smaller one, LeNet, but the patterns are similar across the models. 
Motivated by this observation, we propose a simple yet effective feature selection mechanism, \tool, to selectively drop the identified noisy features, to avoid them passing any information to influence the following decision process. That is, \tool tends to give lower confidence scores to the original high-confidence errors, while the correct predictions are influenced more subtly. As a consequence, the high-confidence errors become more distinguishable from the correct ones, further boosting the fault-revealing capability of simple uncertainty-based methods.

\subsection{Overview of \tool}
\begin{figure}[t]
\centering
\includegraphics[width=0.48\textwidth]{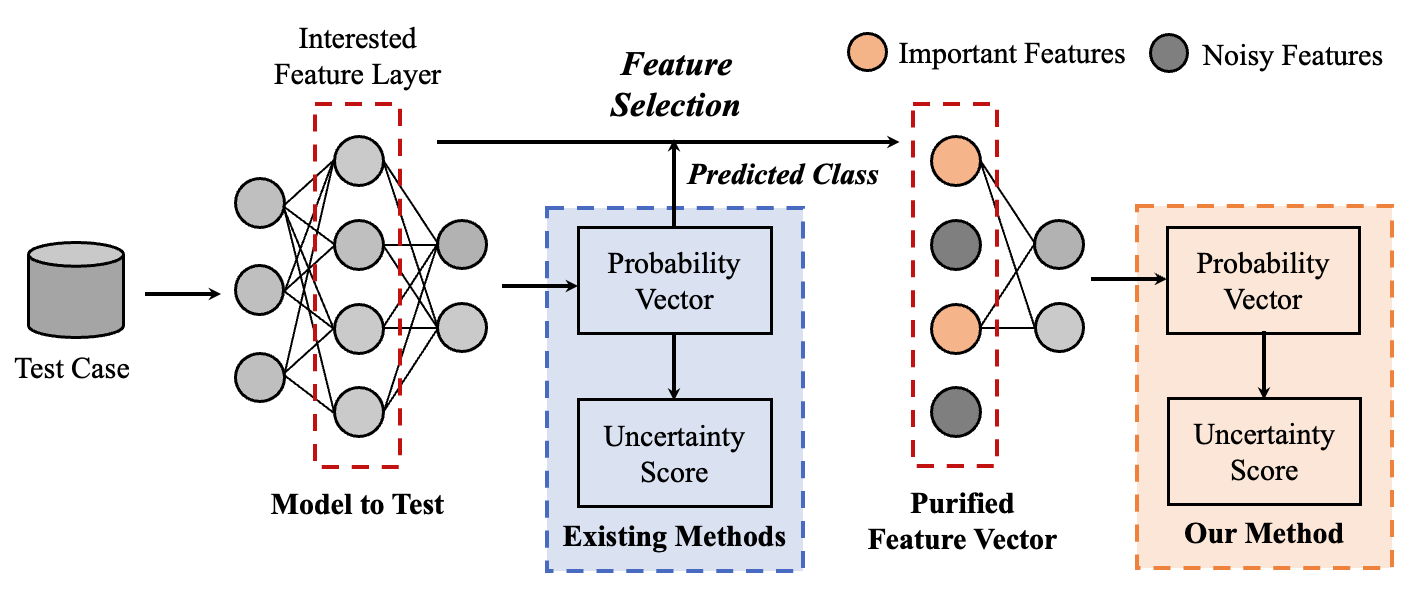}
\vspace*{-4mm}
\caption{An overview of \tool framework. It selectively drops a set of noisy features during inference to derive the probability vector for uncertainty estimation.
}
\label{fig:frame}
\end{figure}

Figure~\ref{fig:frame} presents an overview of the proposed \tool approach, which emphasizes feature selection. Given a model and an unlabeled test suite for prioritization, existing uncertainty-based methods like DeepGini~\cite{deepgini} directly calculate the uncertainty score from the output probability vector, as depicted in the dashed blue box.
In contrast, \tool focuses on performing feature selection at an intermediate layer before calculating the uncertainty score.
Only a part of important features will be passed to the following layers, and other redundant features will be discarded. Then the purified feature vector will be passed to the next layer and derive a new probability vector for the uncertainty calculation, as illustrated in the orange box. Based on the scores, the unlabeled test suite can be ranked for prioritization. Specifically, \tool utilizes an output mask to achieve such a feature selection procedure without the need to modify the model structure or parameters and can be easily applied to any intermediate layer as a plug-in. Note that the additional cost of \tool lies in the propagation of partial layers (usually very few layers), thus only yielding a slight additional run-time cost compared to existing prioritization methods such as DSA~\cite{surprise} and NNS~\cite{nns} that require massive distance computations.

\subsection{Feature Selection}
The feature selection will be performed on a specific hidden layer $l$ of the given DNN. According to the definitions in the background (see Section~\ref{sec:background}), layer $l$ encodes the inputs to a feature space with dimension $N_l$. We denote the feature vector of this layer for a given test input $x$ as $f^l(x)$, which is then passed to the following layers to perform the final prediction. For convolutional layers, each kernel will be considered a basic feature element, while for fully-connected layers, each neuron is the basic element. 
For $C$ classes, we construct a feature mask $M=[M_1,M_2,\cdots,M_C]$ with $N_l\times C$ dimensions, where $C$ is the number of classes and $N_l$ is the number of features. 
$M$ is a binary matrix (only 0 and 1), and $M_c$ identifies the redundant features for each class. Specifically,  the $i^{th}$ element in $M_c$ equal to $0$ indicates the $i^{th}$ feature in $f^l$ is redundant for class $c$ and should be omitted, whereas a value of $1$ indicates that the feature is important and should be preserved.

As illustrated in Fig.~\ref{fig:frame}, based on the originally predicted class $c=\argmax p(x)$, the corresponding mask $M_c$ is then applied to the feature vector $f^l(x)$ to obtain the purified feature vector as:
\begin{equation}
\hat{f}^l(x) = f^l(x) \odot M_c 
\end{equation}
where $\odot$ denotes the element-wise multiplication. 
This operation effectively prevents information from noisy features, which will have consistent zero output values, from being passed to subsequent layers. The modified vector $\hat{f}^l(x)$ is then used to derive the new probability vector $\hat{p}(x)$ through forward propagation by the layers behind $l$. This feature selection process can be easily applied to any hidden layer with the use of an output mask, without the need to modify the model structure.
It is noteworthy that different layers extract different semantic features, where shallow layers tend to extract more common features while deep layers focus on more complex features~\cite{yosinski2015understanding}.  
Based on our empirical studies, we find that the deep layer is a good choice for the feature selection, which effectively enlarges the differences between high-confidence erroneous and correct predictions. We further elaborate on the influence of layer locations in our experiments.

\subsection{Contribution Measurement}
\label{sub:measurement}

To construct the feature mask $M$, we need to quantify the importance (contribution) of each feature for a specific class. The software engineering community has put a significant effort into devising a variety of neuron-level metrics to guide the testing procedure of DNNs in a more targeted way. For instance, DeepInspect~\cite{tian2020testing} utilizes the neuron activation frequency for measuring how the neurons interact with the data from different classes.
The neuron that is more frequently activated (larger than the threshold) is regarded as more important. Moreover, other strategies such as variance-based~\cite{surprise} and gradient-based~\cite{arachne} are also proposed to help measure the importance of neurons. These quantifiers can be naturally used for our purpose. However, due to the highly non-linear interactions of DNNs, it is still unclear how important a neuron (feature) is for the model's final prediction, as a larger output value or frequency does not mean a greater influence on the model's final output.

Instead of relying on existing static internal analyses, we adopt a direct approach inspired by the Shapley Value~\cite{shapley} from the machine learning community to examine the feature importance based on the prediction outcomes. It calculates the overall contribution of neurons by adding them to every possible subnetwork. However, applying such a neuron-wise approach is challenging in our setting, due to its significant computational overhead, which leads to exponential time complexity, making it difficult to meet high-efficiency requirements. Thus, we conduct the testing procedure in a one-by-one manner, similarly to elimination in program analysis, by directly assessing the individual contribution of features to the original model, thereby achieving linear time complexity.
Specifically, for the $i^{th}$ feature $f^l_i$ of layer $l$, we empirically calculate its contribution for a given class $c$ by comparing the model’s output difference with and without feature $f^l_i$ as:
\begin{equation}
\label{v}
    Contribution(f^l_i, D_c)=Conf(f,D_c)-Conf(f^*,D_c)
\end{equation}
where $f$ is the original model and $f^*$ is the model variant where the output of the feature $f^l_i(x)$ is consistently set to zero while keeping the other features unchanged. The function $Conf$ measures the model's average confidence scores over a dataset $D_c$ as:
\begin{equation}
Conf(f,D_c)=\frac{1}{|D_c|}\sum_{x\in D_c}p_c(x) 
\end{equation}
where $p_c$ is the probability score for class $c$, and $D_c$ is a clean dataset consisting of correctly predicted examples with the ground-truth label being class $c$ (we sample from the training dataset). We carefully exclude uncertain examples (with low confidence), thereby focusing on a clear decision-making process and enabling a targeted analysis of features important or redundant for accurate predictions for a class. Specifically, the threshold of $0.9$ is set based on empirical observations with the aim to identify the features important or irrelevant for correctly predicting each class, as high-confidence examples are regarded as representative examples of the class. The intuitive measurement here is that a significant drop in final confidence (a large contribution), when a feature is removed, underscores its importance to the model’s ability to correctly predict class $c$. Note that \tool is flexible and allows integrating other candidate importance measurements to obtain the feature mask, and we elaborate on the impact of different quantification schemes in our experiments.

\begin{algorithm}[t]
    \caption{$\tool(f, f^l, r, T, \lambda)$}
    \label{alg:dd}
    \KwIn{Model to test $f$, feature layer $f^l$, selection rate $r$, unlabeled test suite $T$, uncertainty metric $\lambda$}
    \KwOut{Prioritization order of $T$.}

    Initialize score list $Q$

            \tcp{Preparation Step}
    \For{each class $c\in Y$  }{

        Obtain contribution vector $V_c$ for $f^l$
        
        Obtain binary mask $M_c$ from $V_c$ ($r\%$ of $M_c$ are set $0$)
    }

            \tcp{Prioritization Step}
    \For{$x\in T$}{

        Get original predict label $c$ for $x$
        
        Get original feature vector $f^l(x)$
        
        Get purified feature vector $\hat{f}^l(x) = f^l(x) \odot M_c $
        
        Get probability vector $\hat{p}(x)$ using $\hat{f}^l(x)$
        
        Get uncertainty score $s=\lambda(\hat{p}(x))$

        Append $s$ to list $Q$
    }
    
    \Return $ArgSort(Q)$
\end{algorithm}

After quantifying the contribution of features using Equation~\ref{v}, we get a contribution vector $V_c$ of the features for each class $c$. To construct the binary mask $M_c$, we select the top-k features with the lowest contribution values as redundant, setting their corresponding mask values to $0$ and marking the remaining, more crucial features with $1$.
We define a hyper-parameter $r=\frac{k}{N_l}$ to denote the percentage of features selected to drop. A low $r$ indicates a smaller fraction of redundant features will be pruned and a larger fraction of features are preserved. When $r=0$, the newly derived output $\hat{p}(x)$ becomes the same as the original one $p(x)$. We discuss the influence of the pruning rate later in experiments.

\subsection{\tool for Test Prioritization}
The complete procedure of our \tool method for test case prioritization is detailed in Algorithm~\ref{alg:dd}. \tool is designed to be a generic tool compatible with all existing uncertainty-based methods such as DeepGini~\cite{deepgini}. Given a selected feature layer $f^l$, we first measure the importance of features for each class using a specific quantifier (Line 3), and we then prepare the binary feature mask $M_c$ from $V_c$, setting $r\%$ of it to $0$ (Line 4). During the prioritization process (Line 6-13), for the unlabeled test suite $T$, the objective is to prioritize $T$ such that faults appear at the forefront. For each test case $x\in T$, the predicted label $c$ is first determined (Line 7). The original feature vector $f^l(x)$ is then taken (Line 8). According to the mask $M_c$, only a part of important features are preserved as $\hat{f}^l(x)$ (Line 9). Finally, we get a new probability vector $\hat{p}(x)$ based on $\hat{f}^l(x)$, and we calculate the uncertainty of the test case using the given uncertainty measure $\lambda$ as input (Line 11). Finally, the test samples are ranked by their uncertainty scores (line 14). Compared to traditional uncertainty-based methods, the additional cost with \tool involves constructing the critical feature mask $M$ and executing partial layer propagation  (behind the selected feature layer $f^l$). Since the mask construction is a one-time effort, \tool is considered efficient.

\subsection{Comparison with Similar Work} 
\label{Sec:comp}
Here, we provide a qualitative comparison between \tool and two similar approaches, all aiming to obtain a justified probability vector for distinguishing 
more prediction faults from the correct ones. \tool is similar to Monte-Carlo Dropout (MC-Dropout)~\cite{mcdrop} in its approach of feature dropping in the hidden space. MC-Dropout utilizes the dropout functionality (which is often used in model training) during inference, running the DNN multiple times (e.g., $t$ times) while randomly dropping a proportion of neuron activations. The output probability is then averaged over the $t$ runs. 
In contrast, \tool is guided by a critical feature mask, which helps to drop noisy features in a targeted manner without needing multiple runs for a single test case, thereby achieving much better effectiveness and efficiency across experiments. 
More recently, Nearest Neighbor Smoothing (NNS)~\cite{nns} has been proposed to improve uncertainty-based prioritization methods. NNS adjusts the probability vector by interpolating its original vector with those of its neighbors (typically using K-NN to find the neighbors in the test suite) as:
\begin{equation}
\hat{p}(x)=\alpha \cdot p(x) + \frac{(1-\alpha)}{k} \sum_{x'\in x_{KNN}} p(x')
\end{equation}
where $x'$ denotes the found neighbors, $\alpha$ is the balance factor, $k$ is the number of neighbors, and $\hat{p}(x)$ is the derived new probability output over the original test case and its $k$ neighbors.
However, NNS requires iteratively calculating a heavy distance matrix with the test suite for the K-NN algorithm, which can be particularly costly for large models. In contrast, \tool is more efficient. Once the feature mask is obtained as a one-time effort, it incurs only a slight computational cost during inference. We provide a quantitative comparison of \tool with the two methods in experiments.

\section{Evaluation}
\label{sec:evaluation}

\subsection{Experimental Setup}
\subsubsection{Datasets and Models.} 
Following recent work~\cite{weiss2022simple,nns}, we apply \tool to four benchmark datasets commonly used in the literature, i.e., MNIST \cite{lecun1998gradient}, FASHION \cite{xiao2017fashion}, CIFAR-10 \cite{krizhevsky2009learning}, and SVHN~\cite{netzer2011reading}. For each dataset, we adopt two different DNN model structures for evaluation, where LeNet~\cite{lecun1998gradient}, ResNet~\cite{resnet}, and VGG~\cite{vgg} are standard and popular structures. Conv-6 and Conv-8~\cite{surprise} are the models composed of multiple convolutional layers. Additionally, we extend our evaluation to include three datasets representing high-dimensional image, audio, and text data: Imagenette \cite{Howard_Imagenette_2019}, GSCmd \cite{warden2018speech}, and Imdb \cite{maas2011learning}, respectively.

\subsubsection{Test Suite for Prioritization}
We prepare both clean and noisy test suites to evaluate the test prioritization methods. In the clean setting, we split the original dataset following \cite{nns} and then use the unlabeled clean test samples for evaluation. This is to test how well a prioritization method works on a test suite with the same distribution as the training data and whether it can expose faults earlier. 
Considering that the data processed by the DNN in a practical environment can be diverse, we use corrupted datasets that are manipulated with a range of modifiers inspired by real-world input corruptions such as noise and blur. As discussed in \cite{weiss2022simple,nns}, corrupted data provides model-independent and realistic data compared to adversarial examples. Specifically, we use MNIST-C~\cite{mu2019mnist}, FASHION-C~\cite{weiss2022simple}, and CIFAR-C~\cite{hendrycks2019benchmarking} as provided in the literature. For SVHN, since no corrupted datasets existed, we apply multiple corruption actions following~\cite{hendrycks2019benchmarking} to create an SVHN-C version for a comprehensive evaluation.

\subsubsection{Test Prioritization Baselines.} 
We assess the effectiveness and efficiency of \tool by comparing it with 10 baselines from four different groups, including three coverage-based methods: NAC~\cite{deepxplore}, NBC~\cite{deepgauge}, and OBSAN~\cite{obsan}, an improved version of NBC by incorporating class-wise neuron boundary information; two surprise adequacy (SA)-based methods: DSA~\cite{surprise} and LSA~\cite{surprise}, which also use class-wise information for computation; three uncertainty-based methods: DeepGini~\cite{deepgini}, MaxP~\cite{maxp}, and Margin~\cite{scheffer2001active}; and two post-hoc methods as introduced in Section~\ref{Sec:comp}: MC-Dropout~\cite{mcdrop} and NNS~\cite{nns} (state-of-the-art). 
The uncertainty-based and post-hoc methods are designed to be class-agnostic. We use the same uncertainty calculation method, DeepGini, for the base of evaluated post-hoc methods. We configure each baseline according to the default settings reported in its respective paper. Specifically, for MC-Dropout, we use $t=50$, and for NNS, we set $\alpha=0.5$ and $k=10$. We also consider the pure random prioritization baseline.

\subsubsection{Evaluation Metrics.} 
We use Average Percentage of Fault Detection (APFD)~\cite{apfd} to evaluate the overall performance of a test case prioritization method as:
\begin{equation}
APFD=1-\sum\frac{TF_i}{n*m}+\frac{1}{2*n}
\end{equation}
where $n$ denotes the total number of test cases in the test suite, $m$ denotes  the total number of faults exposed in the DNN under the test suite, and $ TF_i$ denotes the position of the first test in the test suite that exposes fault $i$. The value of APFD ranges from 0 to 1, with higher values representing higher fault detection efficiency.

While APFD is a good measure of the overall performance of test case prioritization methods, it is not suitable for comparing different methods within a given budget, i.e., test case selection based on the prioritization results. Therefore, we also use Test Relative Coverage (TRC)~\cite{nns} to evaluate the performance of a test case prioritization method for a given budget as:
\begin{equation}
TRC=\frac{\#SelectedFaults}{\min(\#SelectBudget, \#TotalFaults)}
\end{equation}

TRC measures how far a prioritization method is from the ideal case in a given budget. The value of TRC ranges from 0 to 1, where a higher value represents a higher proportion of faults exposed within the selection budget. 

\subsubsection{Experimental Environment.} Our experiments were conducted on a computational server with an Intel(R) Xeon(R) CPU E5-2680 v4 @ 2.40GHz, an NVIDIA GTX 2080Ti GPU, and 256GB of RAM. The operating system used was Ubuntu 16.04.5 LTS. We conducted all experiments using Python 3.7.4 and Tensorflow 2.6.0.

\subsection{Research Questions}

In the following, we evaluate \tool through extensive experiments and answer five research questions. 

\begin{table*}[]
   \renewcommand\arraystretch{1.15}
   \setlength\tabcolsep{4.5pt}
   \centering
   \small
   \caption{The APFD values (capability to expose mis-classifications) on clean data. The higher APFD value the better.}
   \vspace*{-2mm}
   \label{tab:clean}
     \begin{tabu}{cc|c|ccc|cc|ccc|ccc} 
\tabucline[1pt]{-}
{\multirow{2}{*}{\textbf{Dataset}}} & \multicolumn{1}{c|}{\multirow{2}{*}{\textbf{Model}}}
& {\multirow{2}{*}{\textbf{Random}}} & \multicolumn{3}{c|}{\textbf{Coverage-based}} & \multicolumn{2}{c|}{\textbf{SA-based}} & \multicolumn{3}{c|}{\textbf{Uncertainty-based}} & \multicolumn{3}{c}{\textbf{Post-Uncertainty}} \\ \cline{4-14} 
&     &     & NAC     & NBC     & OBSAN    & DSA  & LSA  & DeepGini      & MaxP & Margin     & MC-Dropout & NNS & \tool  \\ \tabucline[1pt]{-}
  {\multirow{2}{*}{MNIST}}  & LeNet-5 & 50.64 & 51.86 & 41.62 &  70.41  &  92.06 & 75.97 &  92.88 & 92.91 & 92.78 & 91.83 & 93.54 & \textbf{94.23} \\ 
    & Conv-6 & 52.45 & 40.86 & 46.84 & 88.03 & 93.15 & 78.55 & 94.84 & 94.86 & 94.89 & 94.60 & 95.52 & \textbf{96.37} \\ \tabucline[1pt]{-}
  {\multirow{2}{*}{FASHION}}  & LeNet-5 & 49.73 & 36.57 & 50.03 & 52.79 & 79.44 & 71.33 & 81.24 & 81.46  & 81.49 & 79.31 & 82.03 & \textbf{83.36} \\
    & Conv-8 & 49.20 & 39.95 & 48.89 & 54.53 & 82.01 & 56.26  &  82.45 & 82.48 & 82.42 & 81.65 & 82.79 & \textbf{83.85} \\ \tabucline[1pt]{-}
  {\multirow{2}{*}{CIFAR-10}}   & ResNet-18 & 49.84 & 47.22 & 49.76 & 55.48  & 71.10 & 50.26 &  72.04 & 72.08 & 71.93 & 71.49 & 72.58 & \textbf{75.11} \\ 
    & VGG-16 & 50.65 & 44.48 & 48.69 & 69.89 & 71.74 & 50.01 &  71.07  & 71.00 & 70.92 & 71.28 & 72.26 & \textbf{75.34} \\ \tabucline[1pt]{-}
   {\multirow{2}{*}{SVHN}}   & ResNet-34 & 49.93& 34.21 & 51.06 & 63.12 & 82.60 & 50.75 & 83.84 & 83.96 & 84.04 & 83.48 & 85.60 & \textbf{86.84} \\ 
&  WRN-28  & 51.47 & 41.83 & 48.32 & 66.14 & 82.50 & 52.16 & 85.21 & 85.50 & 85.66 & 84.92 & 86.63 & \textbf{88.30} \\ \tabucline[1pt]{-}
 & \textbf{Average} & 50.49 & 42.12 & 48.15 & 65.05 & 81.83 & 60.66 & 82.95 & 83.03 & 83.02 & 82.32 & 83.87 & \textbf{85.43} \\
    \tabucline[1pt]{-}
\end{tabu}
\end{table*}

\begin{figure*}[]
\centering
\includegraphics[width=0.26\textwidth]{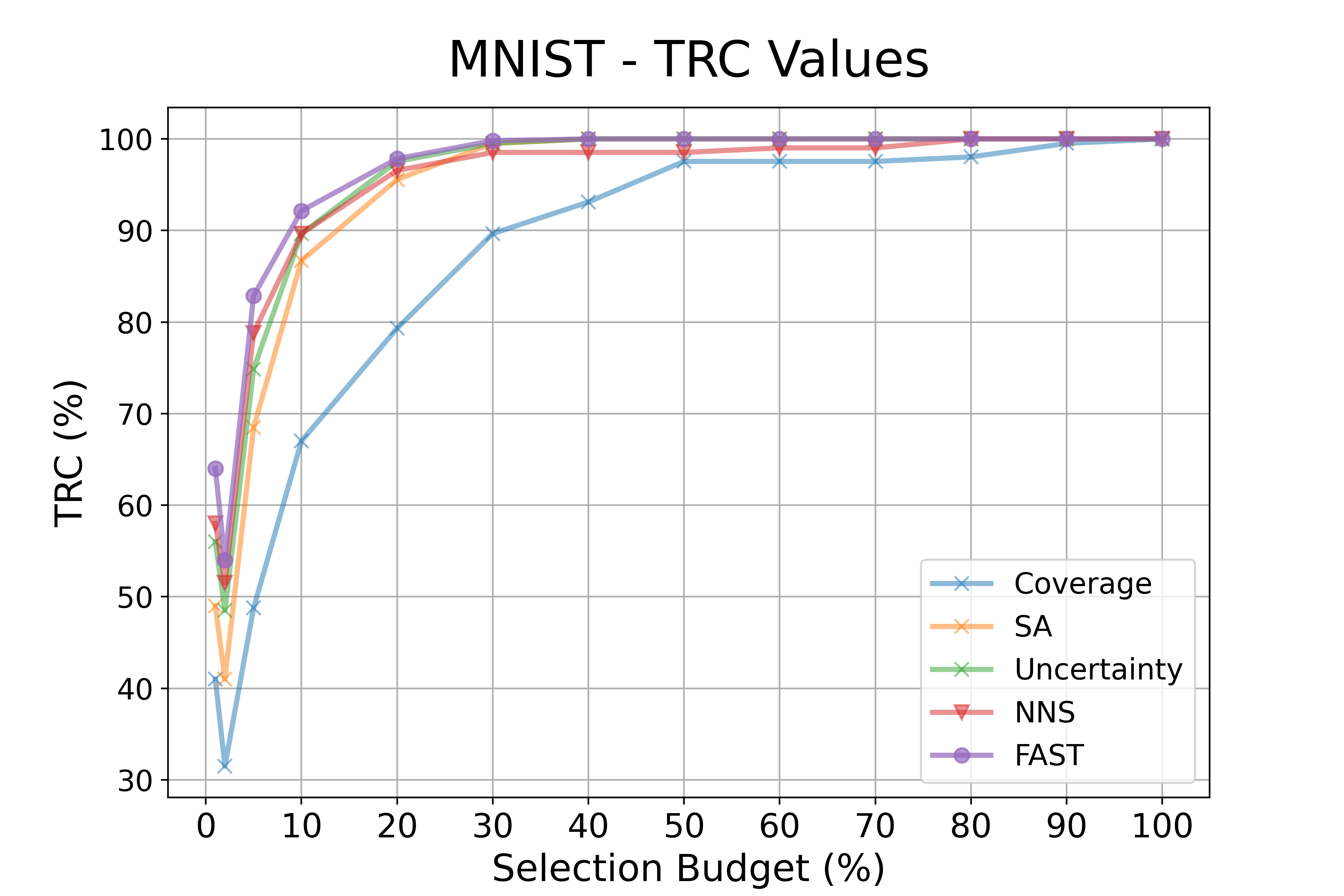}
\hspace{-4mm}
\includegraphics[width=0.26\textwidth]{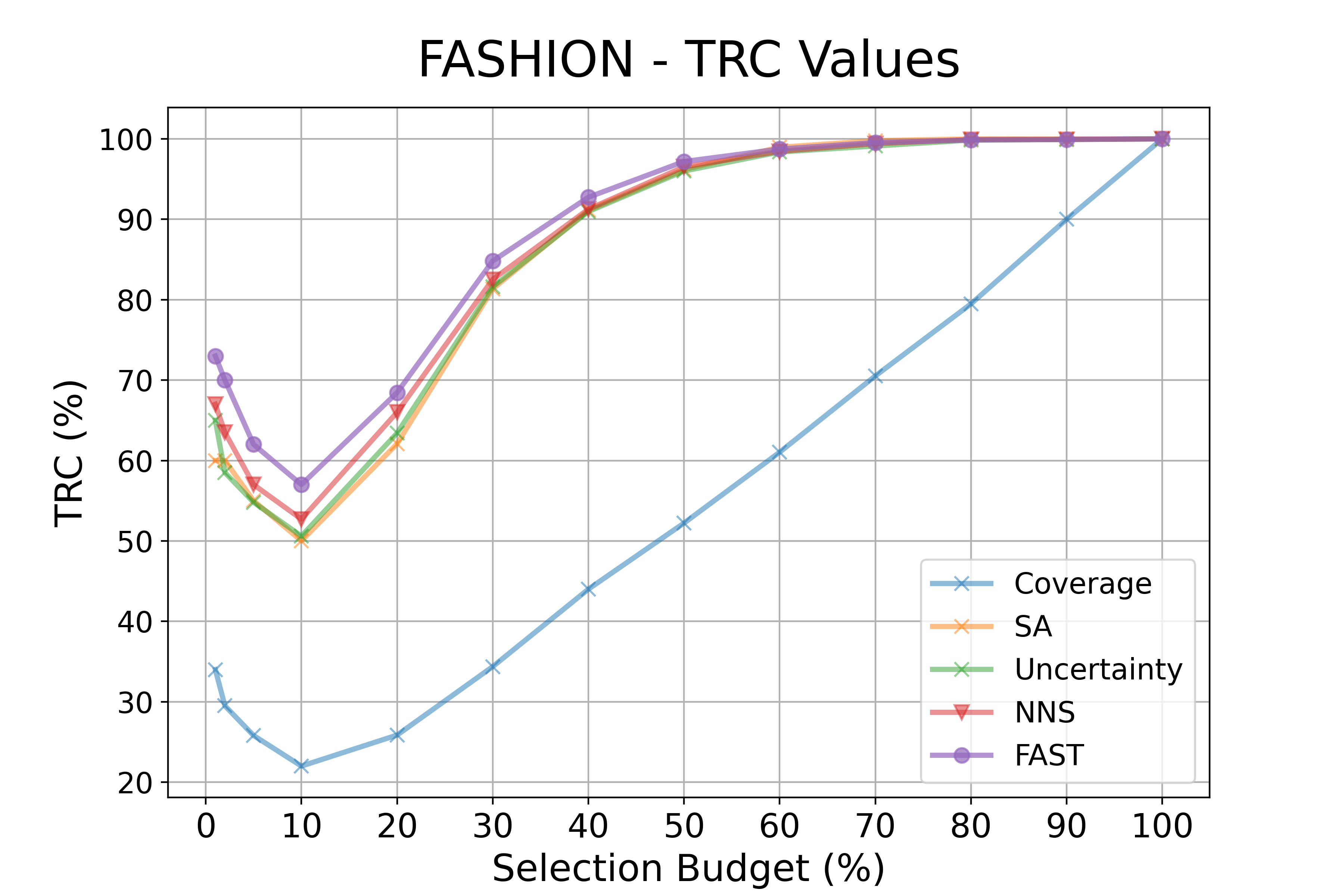}
\hspace{-4mm}
\includegraphics[width=0.26\textwidth]{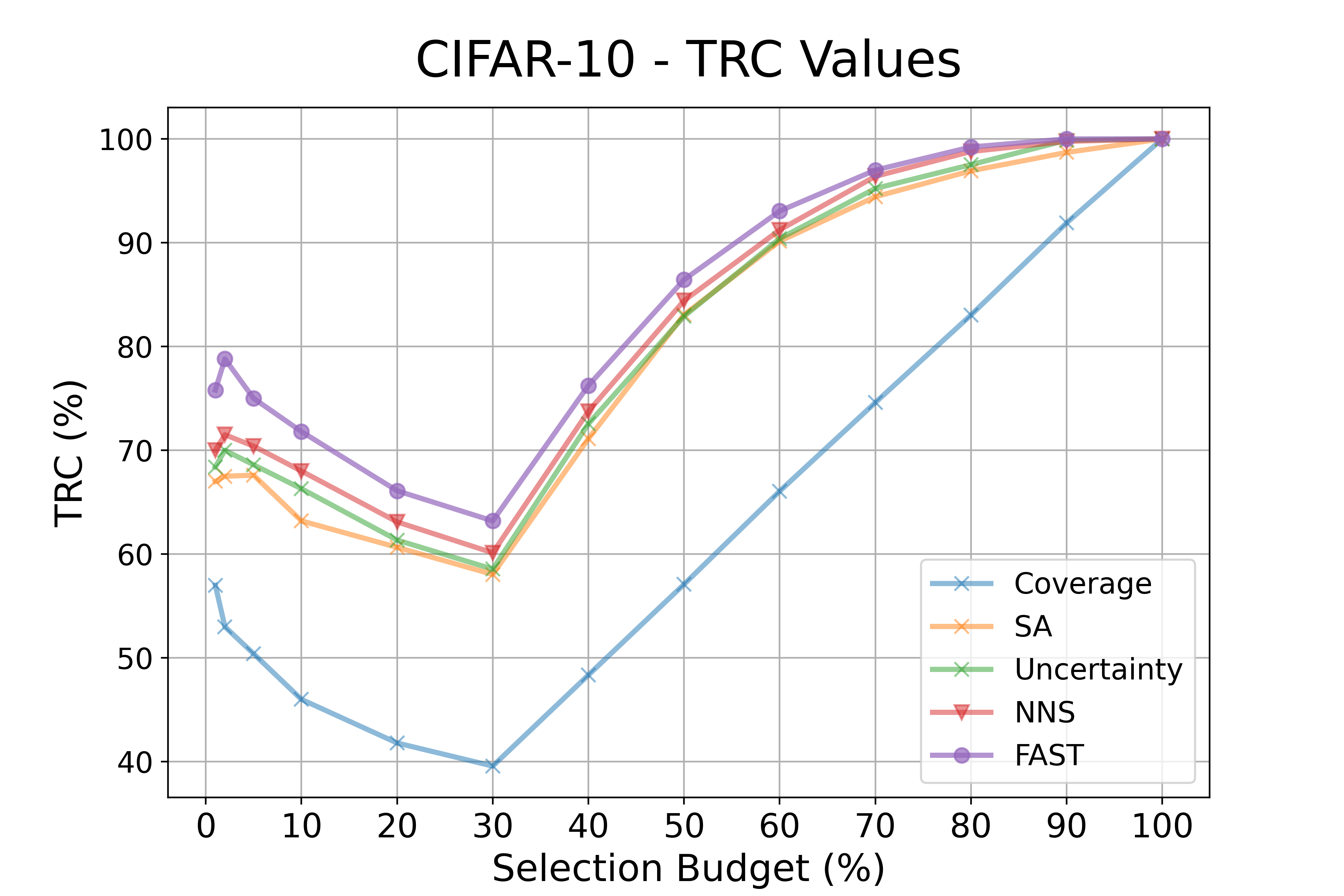}
\hspace{-4mm}
\includegraphics[width=0.26\textwidth]{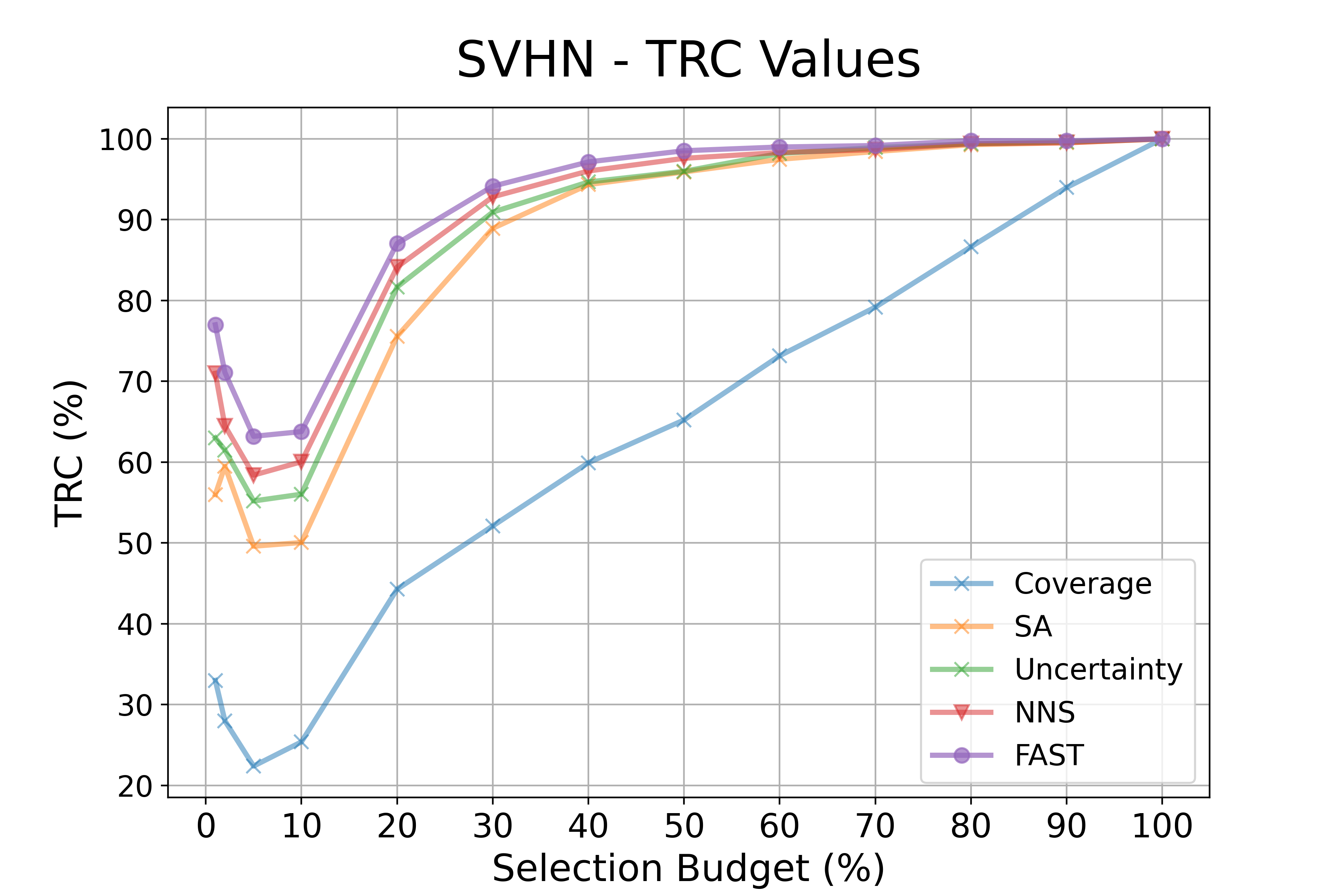}
\caption{The TRC values on clean data. The higher TRC value the better.}
\label{fig:trc}
\end{figure*}

\subsubsection{RQ1: Is \tool more effective in test case prioritization?}\hspace*{\fill}

To answer this question, we evaluate the effectiveness of \tool with representative baselines in prioritizing test cases in the clean and noisy setting, respectively. 

\vspace{0.5mm}
\noindent \textbf{1) Effectiveness on Clean Data}. 
Table~\ref{tab:clean} displays the APFD values obtained by each test prioritization method on clean data, with the highest scores in bold. Across all models tested, \tool achieves the highest APFD values, averaging 85.42\%, which is 3.0\% higher than the uncertainty-based methods (e.g., DeepGini).  This improvement is more than 2.7 times the improvement observed with NNS (about 1.1\%). This suggests the superior capability of \tool in prioritizing misclassifications compared to state-of-the-art methods.

Next, we examine different types of prioritization methods. Generally, uncertainty-based methods outperform coverage-based and surprise-adequacy-based methods, a trend consistent with the findings in the replication study \cite{weiss2022simple}. 
With the APFD values for random ordering around 50\%, two of the three coverage-based methods, i.e., NAC and NBC, perform even worse than the random baseline. This performance can be attributed to the high non-linearity of DNNs, which limits the connections between neuron output magnitudes and prediction faults.
For the SA-based methods, DSA consistently achieves higher values than LSA, still performing slightly worse than DeepGini in 7 out of 8 cases, and about 1.4\% worse on average. LSA shows large fluctuations in performance compared to random, which we attribute to model complexity and the influence of kernel density estimation (KDE) used in LSA, which is heavily affected by the number of features. For instance,  in a hidden layer of the VGG network, which contains thousands of features, the performance of LSA becomes close to random ordering.

\begin{table*}[]
   \renewcommand\arraystretch{1.15}
   \setlength\tabcolsep{4.5pt}
   \centering
   \small
   \caption{The APFD values (capability to expose mis-classifications) on noisy data. The higher APFD value the better.}
   \vspace*{-2mm}
   \label{tab:noisy}
     \begin{tabu}{cc|c|ccc|cc|ccc|ccc} 
\tabucline[1pt]{-}
{\multirow{2}{*}{\textbf{Dataset}}} & \multicolumn{1}{c|}{\multirow{2}{*}{\textbf{Model}}}
& {\multirow{2}{*}{\textbf{Random}}} & \multicolumn{3}{c|}{\textbf{Coverage-based}} & \multicolumn{2}{c|}{\textbf{SA-based}} & \multicolumn{3}{c|}{\textbf{Uncertainty-based}} & \multicolumn{3}{c}{\textbf{Post-Uncertainty}} \\ \cline{4-14} 
&     &     & NAC     & NBC     & OBSAN    & DSA  & LSA  & DeepGini      & MaxP & Margin     & MC-Dropout & NNS & \tool  \\ \tabucline[1pt]{-}
  {\multirow{2}{*}{MNIST}} & LeNet-5 & 49.77 & 54.78 & 41.65  & 61.45 & 83.34  & 67.87  & 83.34 & 83.36 & 83.13 & 82.58 & \textbf{86.15} &  85.88 \\ 
    & Conv-6 & 48.43 & 39.30 & 49.64 & 76.47 & 89.94 & 66.02 & 90.67 & 90.74 & 90.70 & 90.11  & 92.40 & \textbf{93.24}\\ \tabucline[1pt]{-}
  {\multirow{2}{*}{FASHION}} & LeNet-5 & 50.20 & 52.23 & 53.74 & 53.11 & 57.17 & 57.24 & 56.42  & 56.37 & 55.99 & 56.88  & 58.01 & \textbf{58.49} \\
    & Conv-8 & 50.11 & 48.07 & 52.52 & 52.73 & 57.62 & 53.67 & 58.39 & 58.29 & 57.84 & 58.65  & 59.14 & \textbf{60.56} \\ \tabucline[1pt]{-}
   {\multirow{2}{*}{CIFAR-10}}  & ResNet-18 & 50.07  & 47.90 & 50.96 & 51.70  & 60.71 & 51.59 & 61.38  & 61.34 & 61.17 & 61.52 & 61.83 & \textbf{63.85} \\ 
    & VGG-16 & 49.68 & 45.75  & 49.02 & 59.08 & 63.33 & 52.79 &  62.55 & 62.49 & 62.35 &  62.84 & 63.00 & \textbf{64.83} \\ \tabucline[1pt]{-}
 {\multirow{2}{*}{SVHN}} & ResNet-34 & 50.26 & 40.25 & 50.39 & 59.16 & 77.76 & 44.12 & 77.66 & 77.61 & 77.47 & 78.00  & 78.28 & \textbf{80.57} \\ 
 &  WRN-28  & 50.34 & 45.10 & 47.36 & 64.55 & 77.87 & 42.09 & 79.06 & 79.05 & 78.97 & 79.19 & 80.33 & \textbf{81.44} \\  \tabucline[1pt]{-}
& \textbf{Average} & 49.86 & 46.67 & 49.41 & 59.78 & 70.97 & 54.42 & 71.18 & 71.16 & 70.95 & 71.22 & 72.39 & \textbf{73.61} \\ 
    \tabucline[1pt]{-}
\end{tabu}
\end{table*}

\begin{table*}[]
   \renewcommand\arraystretch{1.15}
   \setlength\tabcolsep{3.5pt}
   \centering
   \small
   \caption{Time cost (seconds) for test cases prioritization (per 10K samples).}
   \vspace*{-2mm}
   \label{tab:time}
     \begin{tabu}{cc|ccc|cc|ccc|ccc|cc} 
\tabucline[1pt]{-}
{\multirow{2}{*}{\textbf{Dataset}}} & \multicolumn{1}{c|}{\multirow{2}{*}{\textbf{Model}}}
 & \multicolumn{3}{c|}{\textbf{Coverage-based}} & \multicolumn{2}{c|}{\textbf{SA-based}} & \multicolumn{3}{c|}{\textbf{Uncertainty-based}} & \multicolumn{3}{c|}{\textbf{Post-Uncertainty}} & \multicolumn{2}{c}{\textbf{FAST}}  \\ \cline{3-15} 
&         & NAC     & NBC     & OBSAN    & DSA  & LSA  & DeepGini      & MaxP & Margin     & MC-Dropout & NNS & \tool & S1/S2 & $N_l$   \\ \tabucline[1pt]{-}
   {\multirow{2}{*}{MNIST}}  & LeNet-5  & 20.0s & 33.5s & 34.5s &  916.9s & 180.6s & <5s & <5s & <5s & 872.5s & 228.9s & 21.6s & 16.1s/5.5s  & 84  \\
    & Conv-6  & 21.6s & 40.5s  & 43.8s  & 1108.0s  & 166.2s & <5s & <5s & <5s & 995.1s & 264.3s  & 82.5s  & 64.5s/18.0s & 256 \\ \tabucline[1pt]{-}
   {\multirow{2}{*}{FASHION}} & LeNet-5  & 16.5s &  30.6s & 31.3s & 1015.4s & 322.9s & <5s & <5s & <5s  &  880.4s & 244.5s & 29.3s  &  23.1s/6.2s & 84   \\
    & Conv-8  & 24.9s & 43.2s & 47.3s & 1008.4s & 201.2s  & <5s   & <5s  & <5s  & 916.4s & 288.1s & 35.0s  & 26.2s/8.8s & 128    \\ \tabucline[1pt]{-}
  {\multirow{2}{*}{CIFAR-10}}  & ResNet-18  & 33.8s  & 59.1s &  68.4s  & 1584.2s & 365.8s & <5s  & <5s & <5s & 1133.8s & 395.9s & 126.2s  & 99.0s/27.2s & 512   \\ 
    & VGG-16  & 65.2s & 119.2s & 130.7s & 1985.1s & 620.8s  &  <5s   & <5s  & <5s & 1833.7s & 642.8s & 306.1s  & 273.0s/33.1s & 4096  \\ \tabucline[1pt]{-} 
  {\multirow{2}{*}{SVHN}} & ResNet-34  & 37.5s & 55.6s & 70.2s & 1730.5s & 450.7s & <5s   & <5s  & <5s& 1212.6s & 422.6s & 174.3s  & 134.6s/39.7s & 512   \\ 
 & WRN-28  & 41.4s & 67.1s & 80.5s & 1331.2s & 329.7s  & <5s   & <5s  & <5s  & 1306.3s & 455.3s & 203.2s  & 155.7s/47.5s & 640   \\ \tabucline[1pt]{-}
& \textbf{Average} & 32.6s & 56.1s & 63.3s & 1335.0s & 329.7s  & <5s & <5s & <5s & 1143.9s & 367.8s & 122.3s & 99.0s/23.3s & --   \\  \tabucline[1pt]{-}
\end{tabu}
\end{table*}

As expected, the three uncertainty-based methods, DeepGini, MaxP, and Margin, exhibit similar performance, with their average APFD values all around 83\%. Interestingly, the MC-Dropout method, a post-hoc baseline that averages multiple predictions through random dropout, causes a slight decline (about 0.75\%) compared to the original uncertainty-based methods. This drop is likely because MC-Dropout occasionally drops some key features, which can confuse the decision process. In contrast, \tool, by selectively dropping features guided by contribution measurements, achieves stable improvements over uncertainty-based methods and consistently outperforms them. Although the improvement of NNS over DeepGini is noticeable, it is sometimes subtle when compared to \tool. We hypothesize that NNS tends to select neighboring test cases with similar feature patterns, resulting in outputs that are too similar, which minimally influences the averaged probability outputs, and the difference between errors and correct
may not be significantly enlarged in certain tasks.

While the APFD value serves as an effective overall performance metric, it does not precisely reveal how many errors can be exposed within a given selection budget. Fig.~\ref{fig:trc} shows the TRC scores obtained by each method under various budgets, ranging from 1\% to 100\% of the test suite. As TRC reaches 100\% when the entire test suite is selected, it becomes less meaningful for detailed comparisons. Our focus is therefore on the TRC values at smaller budgets, where a higher TRC score indicates that more faults can be uncovered at a lower cost. The TRC curves for each method exhibit a distinctive turning point; before this turning point, differences between methods are more pronounced but diminish as the selection budget increases. \tool consistently achieves higher TRC values than all other baselines across the selection budget levels, indicating that it can expose more errors with the same budget compared to other methods. This advantage is particularly notable at small selection budgets. To quantify this, we calculated the average TRC values before the turning point for each method. \tool shows substantial improvements of 12.92\%, 14.46\%, 9.54\%, and 17.58\% over the uncertainty-based methods, and 7.76\%, 9.08\%, 6.84\%, and 8.97\% higher than NNS across the four datasets, respectively. These results highlight a significant enhancement in the practical application of \tool for test case selection compared to existing methods.

\noindent \textbf{2) Effectiveness on Noisy Data}. 
Table~\ref{tab:noisy} shows a comparison of the APFD values obtained by each method on noisy data. 
The overall trend is similar to what we observe with clean data, though all evaluated methods show a certain drop in APFD when compared to their performance on clean data (Table~\ref{tab:clean}), indicating that noisy data is generally harder for test case prioritization. Nevertheless, \tool achieves the highest APFD values in 7 out of 8 cases, with the only exception being the MNIST dataset,  where NNS surpasses \tool. The improvement of \tool over the uncertainty-based methods is more than 3\% on average. Additionally, we note that DSA surpasses the uncertainty-based methods in 3 cases, and the improvement of NNS on noisy data is more pronounced than on clean data (about 1.7\%), indicating a noticeable advantage when dealing with data from different distributions.

\begin{tcolorbox}[fonttitle = \bfseries, boxsep=3pt, left=6pt, right=6pt, top=2pt,bottom=2pt]
  \textbf{Answer to RQ1:} \tool consistently improves the performance of uncertainty-based test case selection methods and outperforms the state-of-the-art.
\end{tcolorbox}

\subsubsection{RQ2: Is \tool more efficient?}\hspace*{\fill}
\label{rq2}

To answer this question, we report the total time cost of each method for prioritizing the same test suite (10K samples) in Table~\ref{tab:time}, inclusive of any necessary preparation time.  The time cost for \tool consists of two parts, the feature mask preparation ($S1$) and running FAST during inference for prioritization ($S2$). We include the number of features $N_l$ and split the overall time cost of FAST into two parts as $S1/S2$. 

\tool demonstrates  marked efficiency in prioritizing test inputs, requiring less than 2 minutes on average across various models. This efficiency is over 3 times and 9 times greater than that of other post-hoc methods, NNS and MC-Dropout, respectively. MC-Dropout involves randomly sampling multiple stochastic predictions for the same test input, while NNS requires computing the heavy distance matrix between the test sample and the entire dataset to find the neighbors, both resulting in significantly higher computational costs compared to \tool. The primary cost for \tool is the feature contribution estimation step (for obtaining mask $M$), which varies depending on the target model structure, taking from 20 seconds for smaller models to about 5 minutes for larger models such as VGG-16. Generally, the feature contribution assessment ($S1$) comprises the majority of the overall time cost. It is noteworthy that the estimation step of \tool is a one-time effort, after which the additional inference cost for \tool involves merely the information propagation of few layers (influenced by the layer location). The time overhead incurred for the contribution assessment mainly depends on the number of features $N_l$ of the selected layer. As the assessment procedure for each feature is independent, it can be further accelerated using parallel processing.

In comparison across different method types, undoubtedly, pure uncertainty-based methods emerge as the most efficient, requiring less than 5 seconds (with GPU acceleration), as they do not require any preparatory work. The time cost here reflects the actual time needed for the DNN to make the final predictions. Coverage-based methods incur considerably higher costs than uncertainty-based methods, as they need to extract fine-grained information from the internals and maintain a heavy neuron map, yet their performance does not justify these costs. SA-based methods take even longer than coverage-based methods. Specifically, DSA is the most time-consuming among the evaluated methods as it involves traversing all training data to identify the closest neighbor sample. 

\begin{tcolorbox}[fonttitle = \bfseries, boxsep=3pt, left=6pt, right=6pt, top=4pt,bottom=4pt]
  \textbf{Answer to RQ2:} \tool is more efficient than the SA-based prioritization methods and other post-hoc methods. 
\end{tcolorbox}

\begin{figure*}[]
\centering
\includegraphics[width=0.24\textwidth]{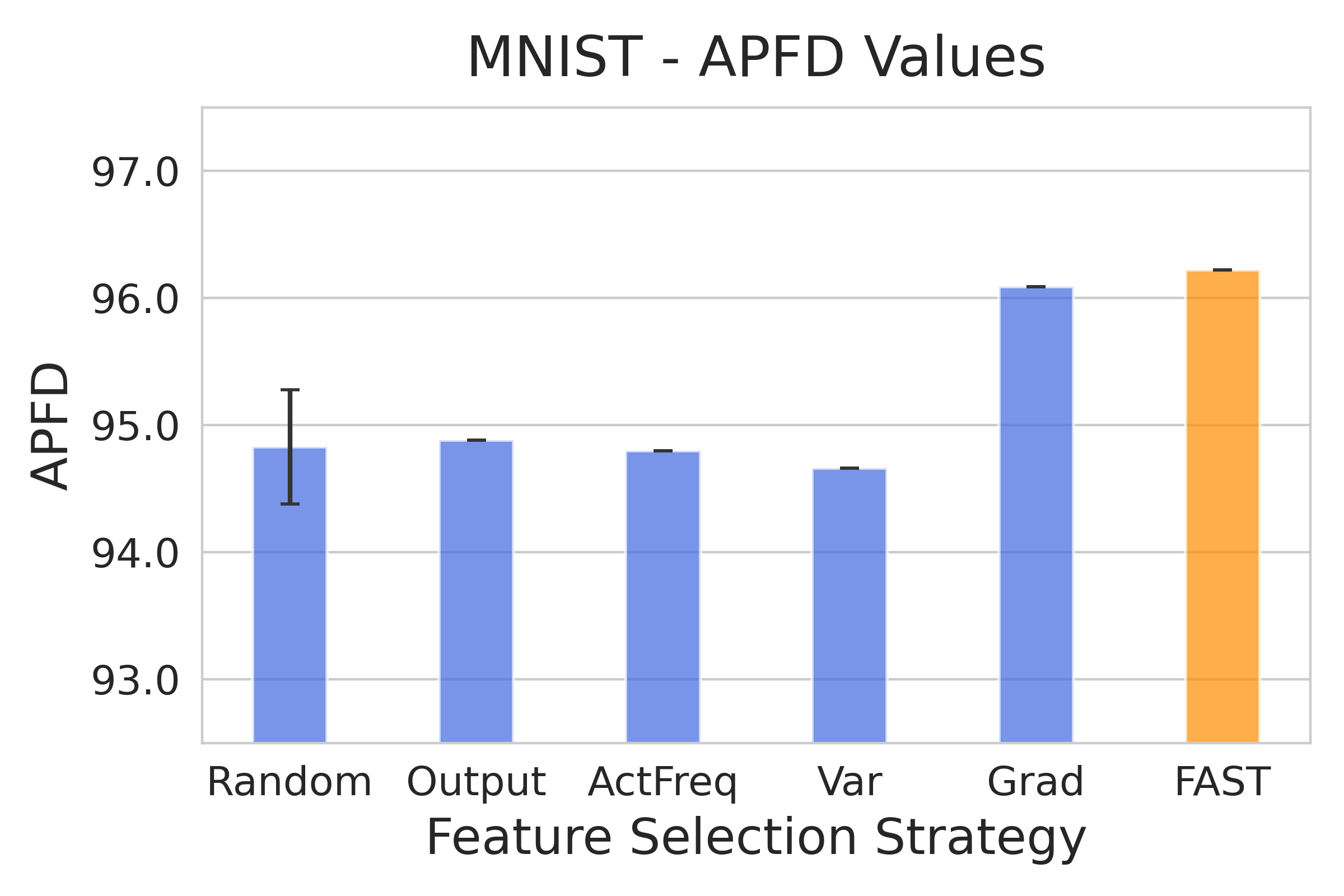}
\includegraphics[width=0.24\textwidth]{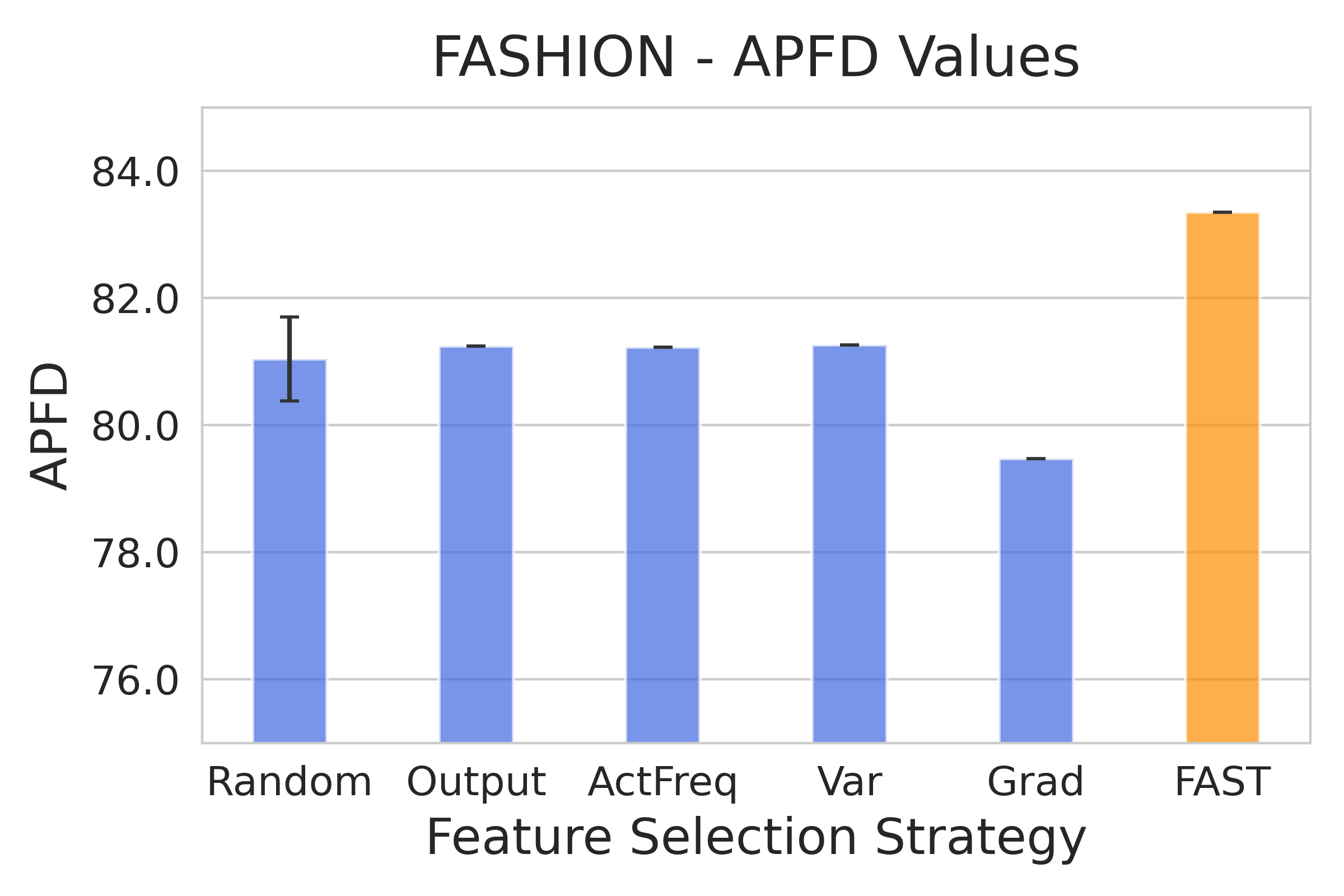}
\includegraphics[width=0.24\textwidth]{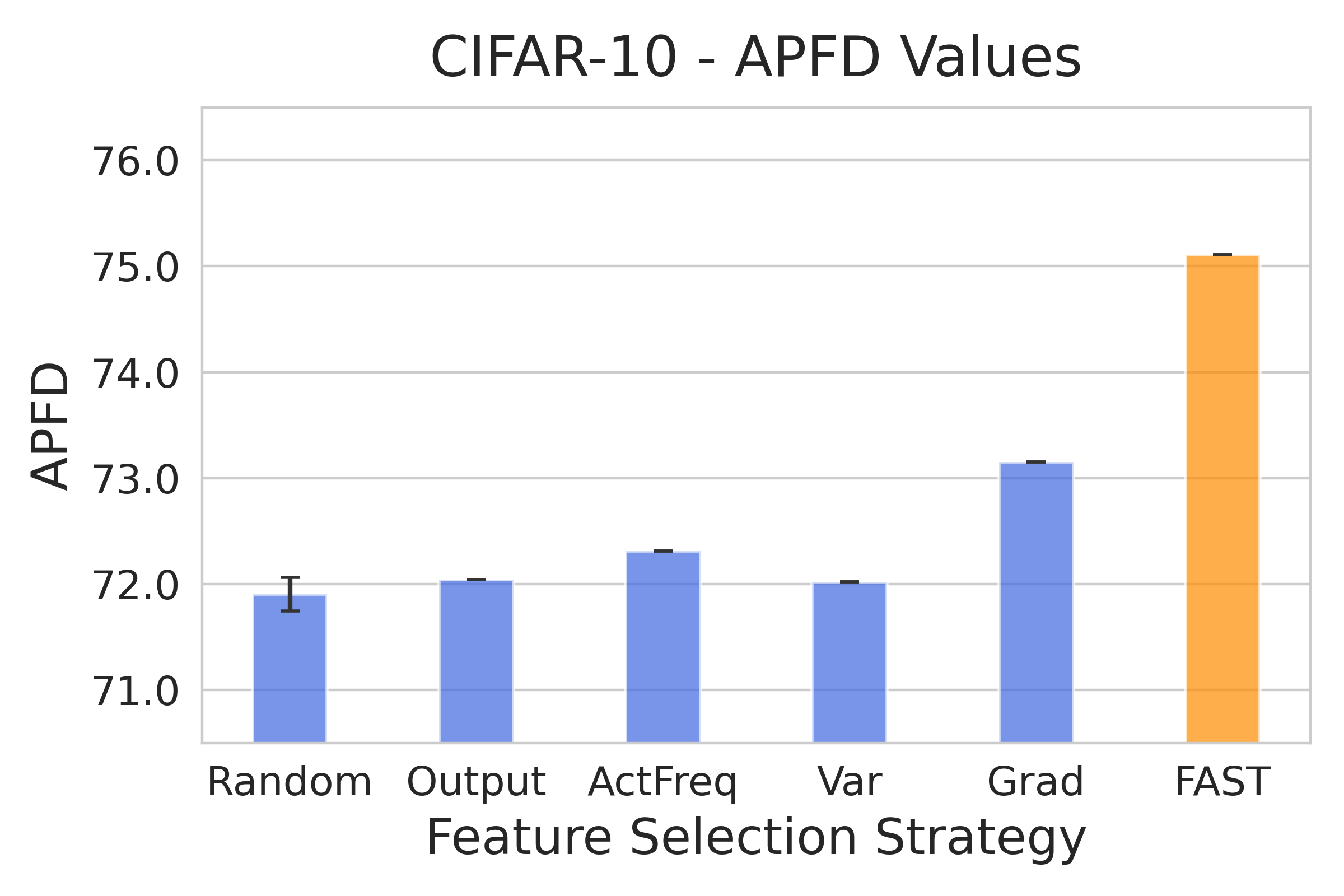}
\includegraphics[width=0.24\textwidth]{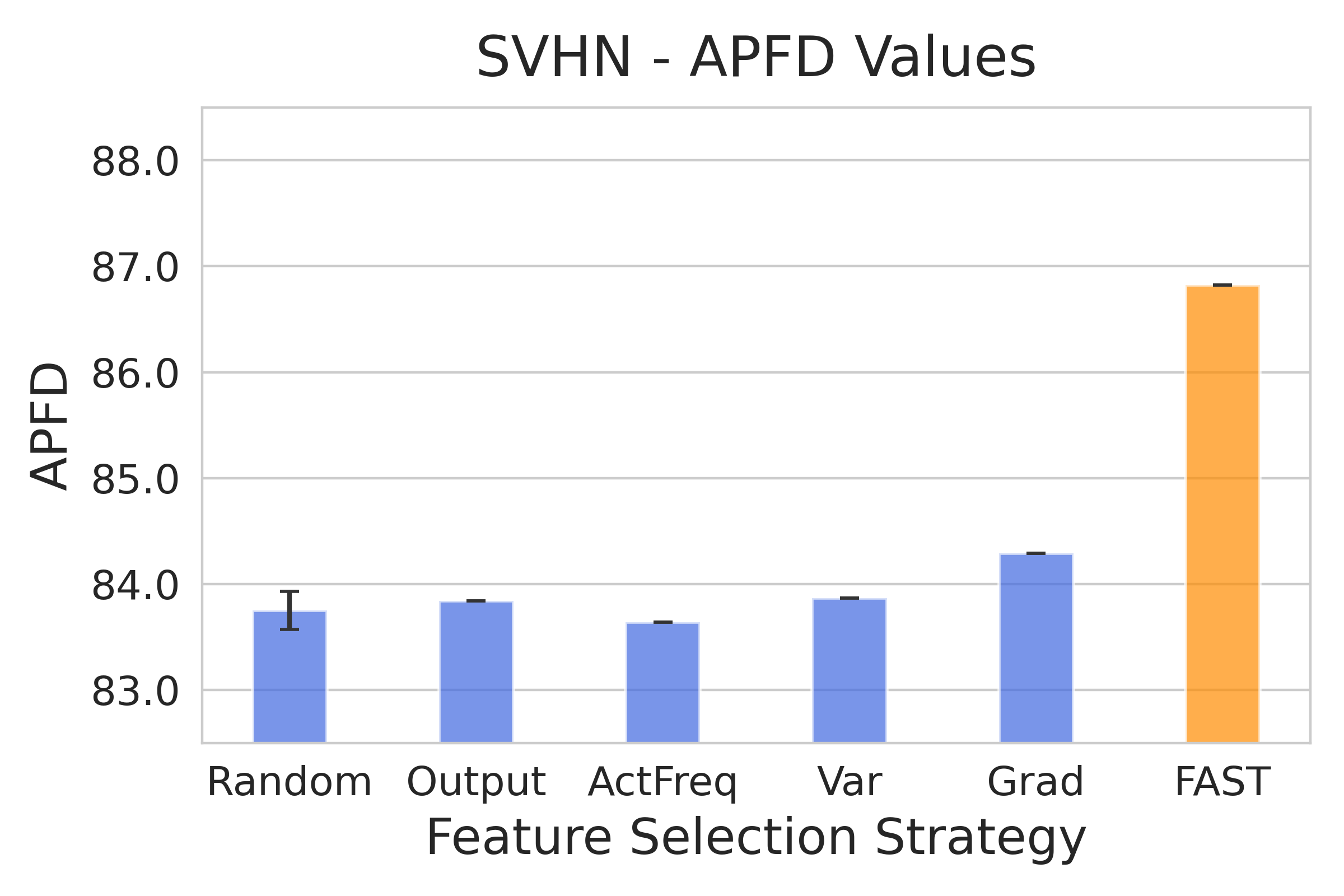}
\caption{The performance of \tool with different feature selection strategies.}
\label{fig:strategy}
\end{figure*}

\begin{figure*}[]
\centering
\includegraphics[width=0.24\textwidth]{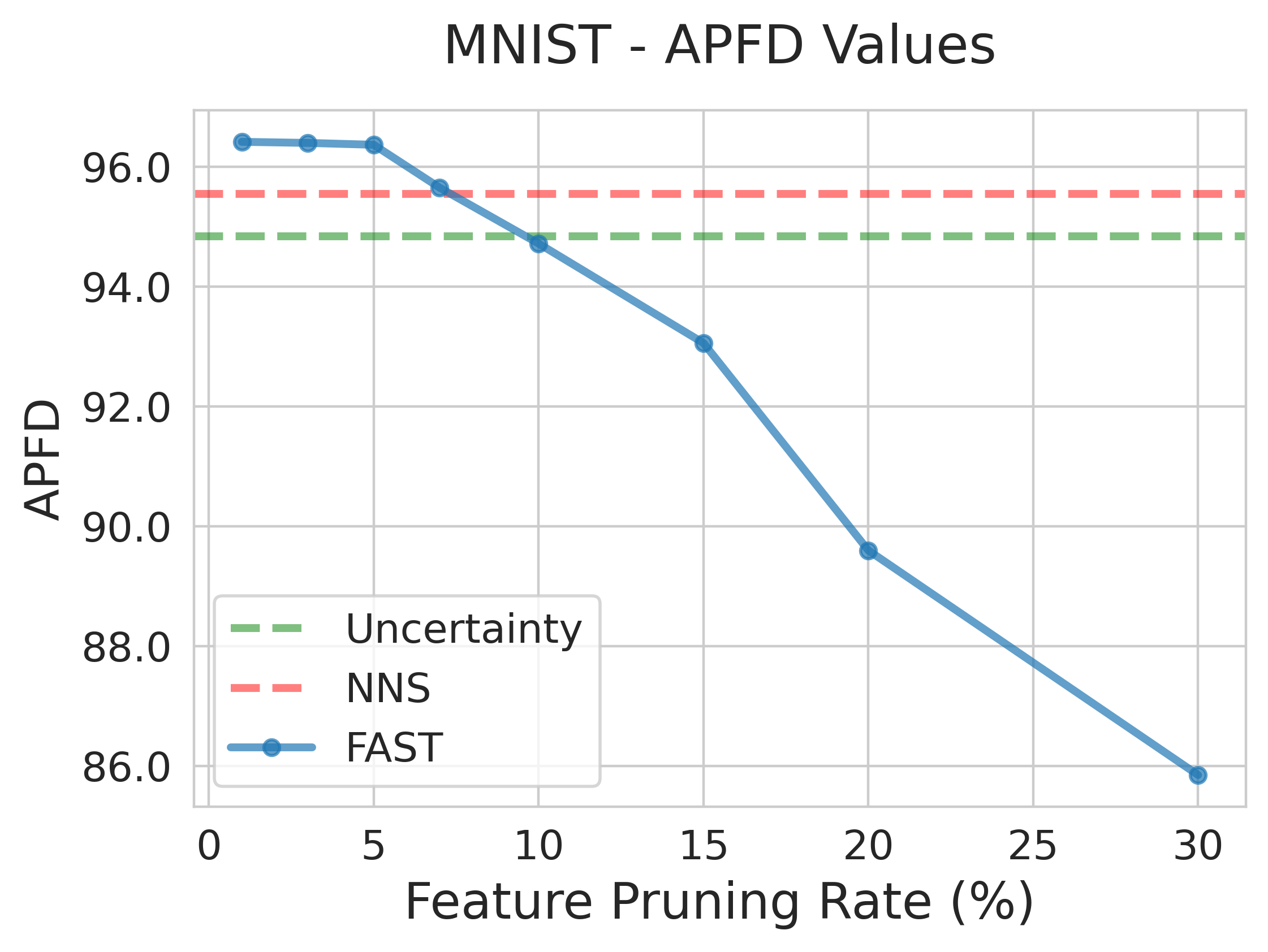}
\includegraphics[width=0.24\textwidth]{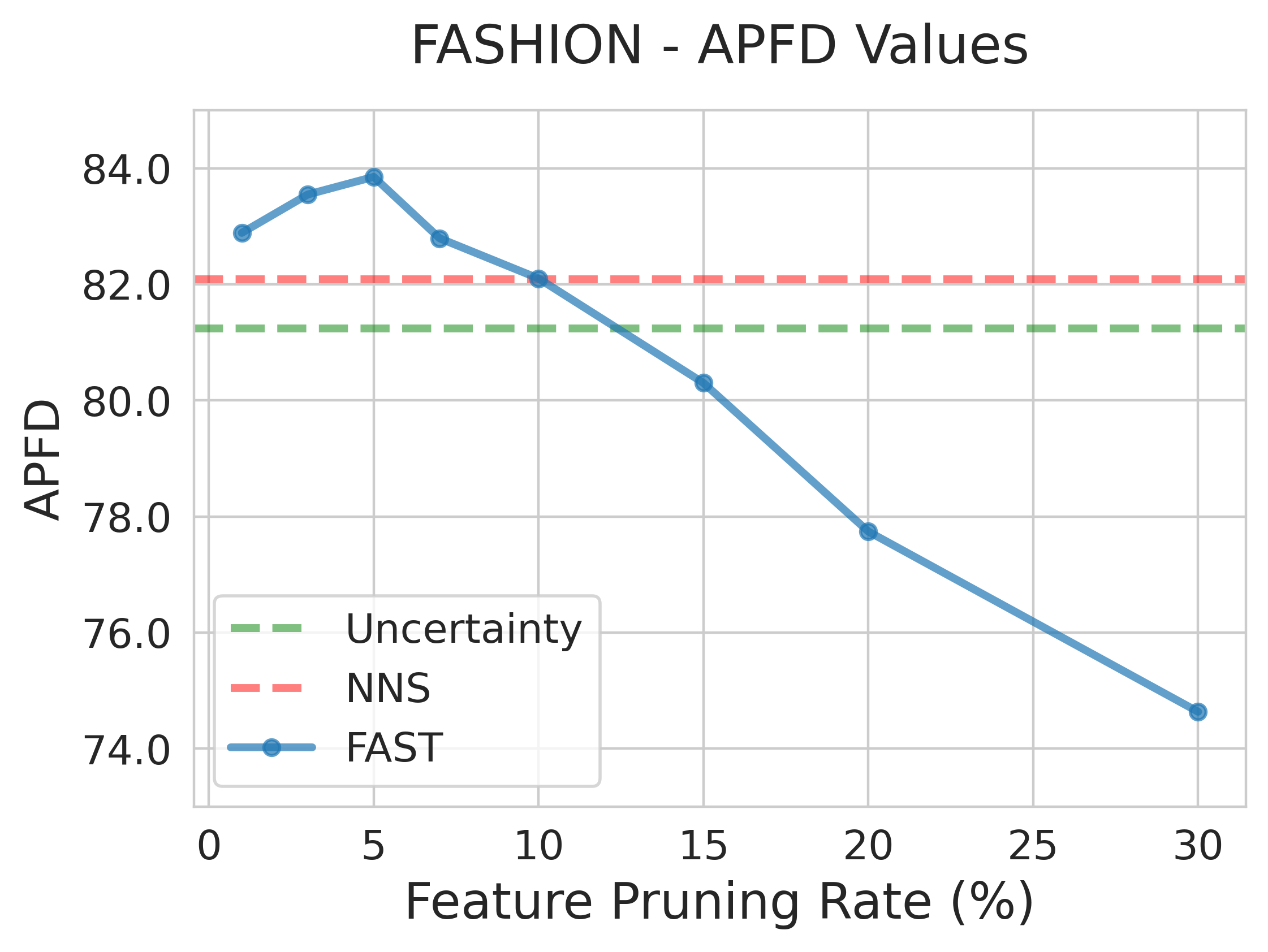}
\includegraphics[width=0.24\textwidth]{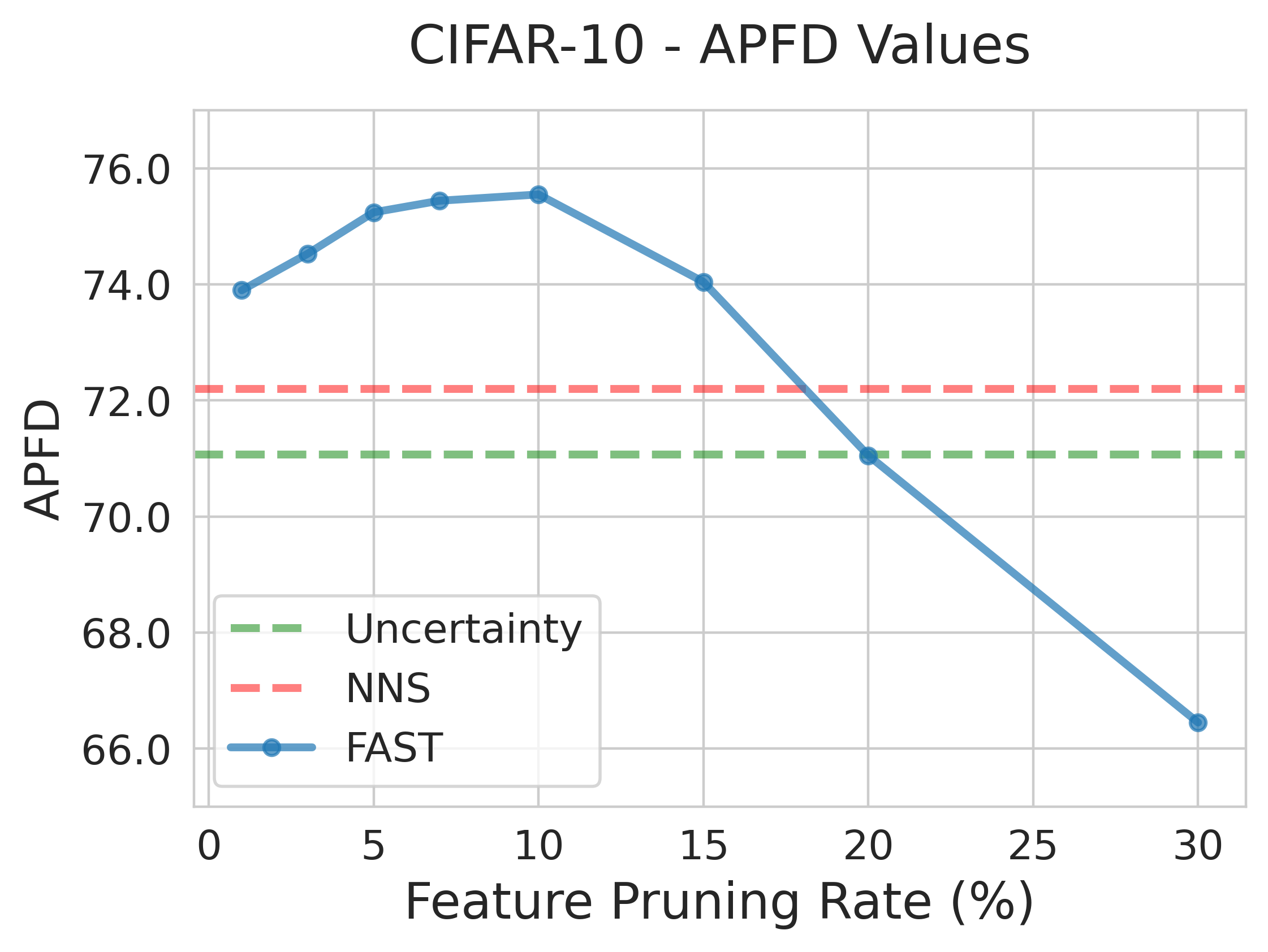}
\includegraphics[width=0.24\textwidth]{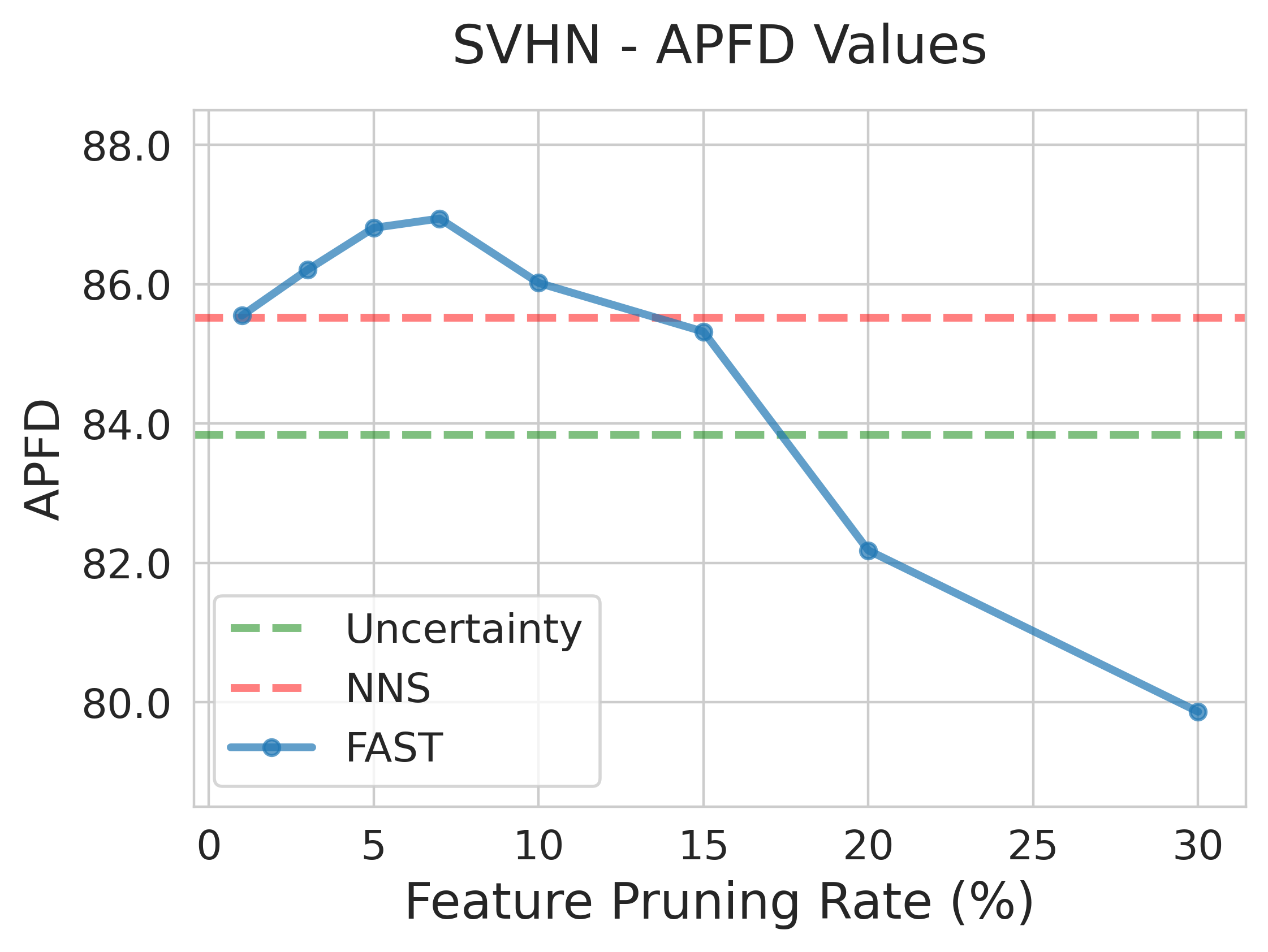}
\caption{The performance of \tool with different feature selection rates.}
\label{fig:rate}
\end{figure*}

\subsubsection{RQ3: What factors influence the performance of \tool?}\hspace*{\fill}
\label{rq3}

To understand how \tool performs under different settings, we conduct an ablation study for \tool on three main influencing factors: the feature selection strategy, the feature selection (pruning) rate, and the layer location where \tool is applied. 

\noindent \textbf{1) Feature Selection Strategy}. 
The core of our feature selection procedure is the importance quantification strategy of each feature regarding the predictions (as detailed in Section~\ref{sub:measurement}). Here, we replace the measurement method used in \tool with five representative strategies popular in the SE community. The Output strategy~\cite{deepxplore} supposes that features with higher output values have  stronger influence on the model decision, and thus will prune the features with the lowest average output values across the validation dataset. The Activation Frequency strategy~\cite{tian2020testing} quantifies how frequently a neuron (feature) is activated (larger than a threshold) during inference, considering rarely activated features as unimportant. The Variance strategy~\cite{surprise} assumes that features with lower variance contain less information, thus contributing less to the predictions. The Gradient strategy~\cite{arachne} calculates the gradients of the model predictions with respect to the feature outputs to measure the contribution. We also consider the Random strategy, where features will be selected and pruned randomly. For consistency, we set the same feature pruning rate ($r=5\%$) for all methods.

Fig.~\ref{fig:strategy} shows the results of \tool using different feature selection strategies. Notably, our method (represented by orange bars) demonstrates significantly higher APFD values compared to all baseline strategies, indicating the pruned features have stronger connections with the prediction errors. This distinction allows errors to become more distinguishable from correct predictions after pruning the identified noisy features. The performance of baseline methods varies considerably. For instance, the Gradient-based strategy achieves comparable performance to ours on the MNIST dataset but performs the worst on the FASHION dataset.  In some cases, other strategies even underperform the random baseline. This suggests that the features selected by these baseline strategies have limited influence on the model's final predictions. We speculate that these methods quantify the features based on  static outputs (except for the gradient-based, which incorporates the model's outputs in its calculation), often assuming that higher activation values or frequencies indicate greater contributions to final decisions. However, this may not hold true due to the high non-linearity of DNNs, which involves complex layer-by-layer propagation. 
In contrast, our method adopts a more intuitive approach by dynamically pruning and directly testing the contribution of each feature with respect to the observable prediction results, which leads to better performance in the feature selection step across the tested models.
Moreover, experiments in RQ1 have already validated that metrics based on output values (e.g., NAC and NBC) have limited or even negative correlations with DNN faults. 
The results underline the superiority of our contribution-based method over static statistical methods in feature selection for complex DNNs. Although our method requires a higher computational cost than the baselines, it plays an indispensable role in  the effectiveness of \tool.

\begin{figure}[]
\centering
\includegraphics[width=0.43\textwidth]{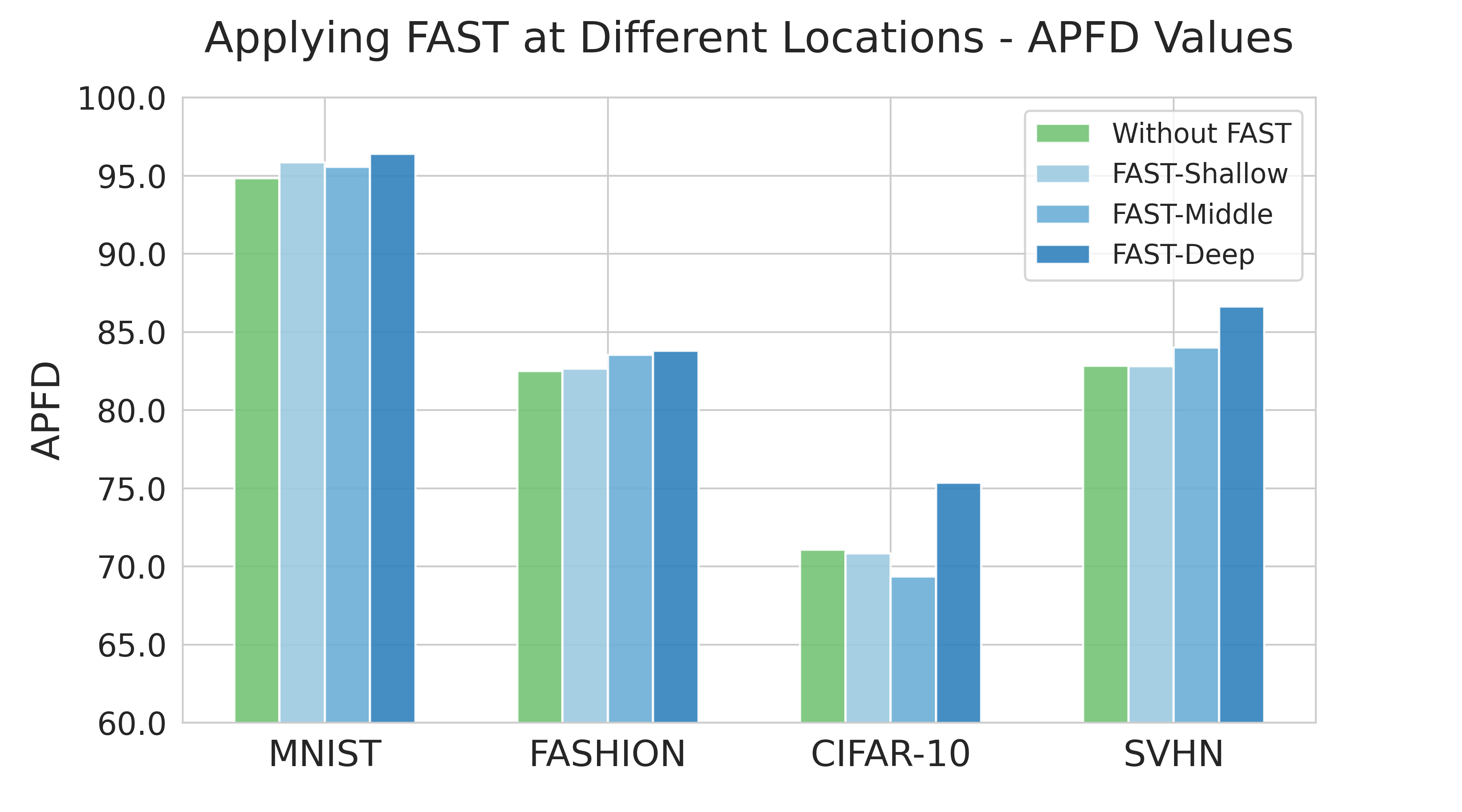}
\vspace*{-1mm}
\caption{The performance of \tool with different layers.}
\label{fig:layers}
\end{figure}

\noindent \textbf{2) Feature Selection Rate}. 
Next, we vary the feature selection (pruning) rate and evaluate the corresponding effects of \tool on the APFD values. We also present the results of baseline uncertainty-based methods, i.e., DeepGini and NNS, for a better comparison.

Fig.~\ref{fig:rate} summarizes the trends of \tool with different pruning rates. We observe that when $r$ is relatively small (below 5\%), the APFD score increases steadily as $r$ increases across datasets (MNIST is an exception that achieves optimal performance at an early stage). This finding aligns with the insights from Fig.~\ref{fig:featurecontribution} that there is a portion of noisy features contributing to the high-confidence errors. As more of the noisy information is discarded, the separability between correct and erroneous predictions increases. However, as $r$ continues to increase, the APFD values drop significantly, suggesting a trade-off between the selection rate and the effectiveness of \tool. We suppose that pruning a larger portion of features will inevitably block some essential information influencing the decisions on correct examples, leading to decreased confidence in these classifications and making it harder to distinguish the errors from correct predictions. 
Note that the optimal $r$ appears to vary depending on the specific characteristics of the model. Based on the visualized results, a rate between 3\% to 10\% is generally suitable for \tool, consistently yielding higher APFD values than the baselines.

\begin{table}[]
\renewcommand\arraystretch{1.15}
   \setlength\tabcolsep{5pt}
   \centering
   \small
    \caption{The APFD values on clean data.}
    \vspace*{-1mm}
    \label{tab:scala}
\begin{tabu}{cc|cccc} \tabucline[1pt]{-}
\textbf{Dataset}    & \textbf{Model}   & \textbf{Random} & \textbf{DeepGini} & \textbf{NNS} & \textbf{FAST} \\ \tabucline[1pt]{-}
Imdb       & LSTM-4 &  49.74     &  69.65   & 69.69    & \textbf{70.47}     \\  
GSCmd & LSTM-6 &   50.53       &  88.49         &  88.50   &  \textbf{89.40} \\
Imagenette & VGG-16  & 49.82       &  65.85     &  66.41   &  \textbf{67.53}   \\ \tabucline[1pt]{-}
\end{tabu} 
\end{table}

\noindent \textbf{3) Feature Selection Layer}. 
We further explore the impact of applying \tool to different layer locations within the model, specifically examining the shallow, middle, and deep layers. We divide the model into three chunks, each consisting of $1/3$ of the total number of layers, and we select one layer from each block to apply FAST. Specifically, the selected layer indices are $1/3/5$, $1/4/7$, $2/8/14$, $4/16/32$ for the Conv-6, Conv-8, VGG-16, ResNet-34 models, respectively. We also present the results without \tool (i.e., pure uncertainty calculation) for a better comparison. 

Fig.~\ref{fig:layers} shows the results. It is evident that applying \tool at the deep layer consistently achieves the best performance across all datasets. Conversely, when \tool is applied to the shallower layers, it sometimes underperforms compared to the baseline without using \tool. This indicates that the feature selection in these layers can sometimes hurt the model's normal decision-making process for correct predictions. This observation aligns with the understanding that shallow layers tend to extract more abstract features, while deeper layers encode higher-level, more complex features~\cite{yosinski2015understanding}, which contain richer information for \tool to distinguish the errors more effectively. Therefore, the deep layer is the preferred location for applying \tool to optimize performance.

\begin{tcolorbox}[fonttitle = \bfseries, boxsep=3pt, left=6pt, right=6pt, top=4pt,bottom=4pt]
  \textbf{Answer to RQ3:} \tool's performance can be influenced by several factors. Configuring a suitable setting is important for \tool to achieve satisfying performance. 
\end{tcolorbox}

\begin{figure*}[]
\centering
\includegraphics[width=0.23\textwidth]{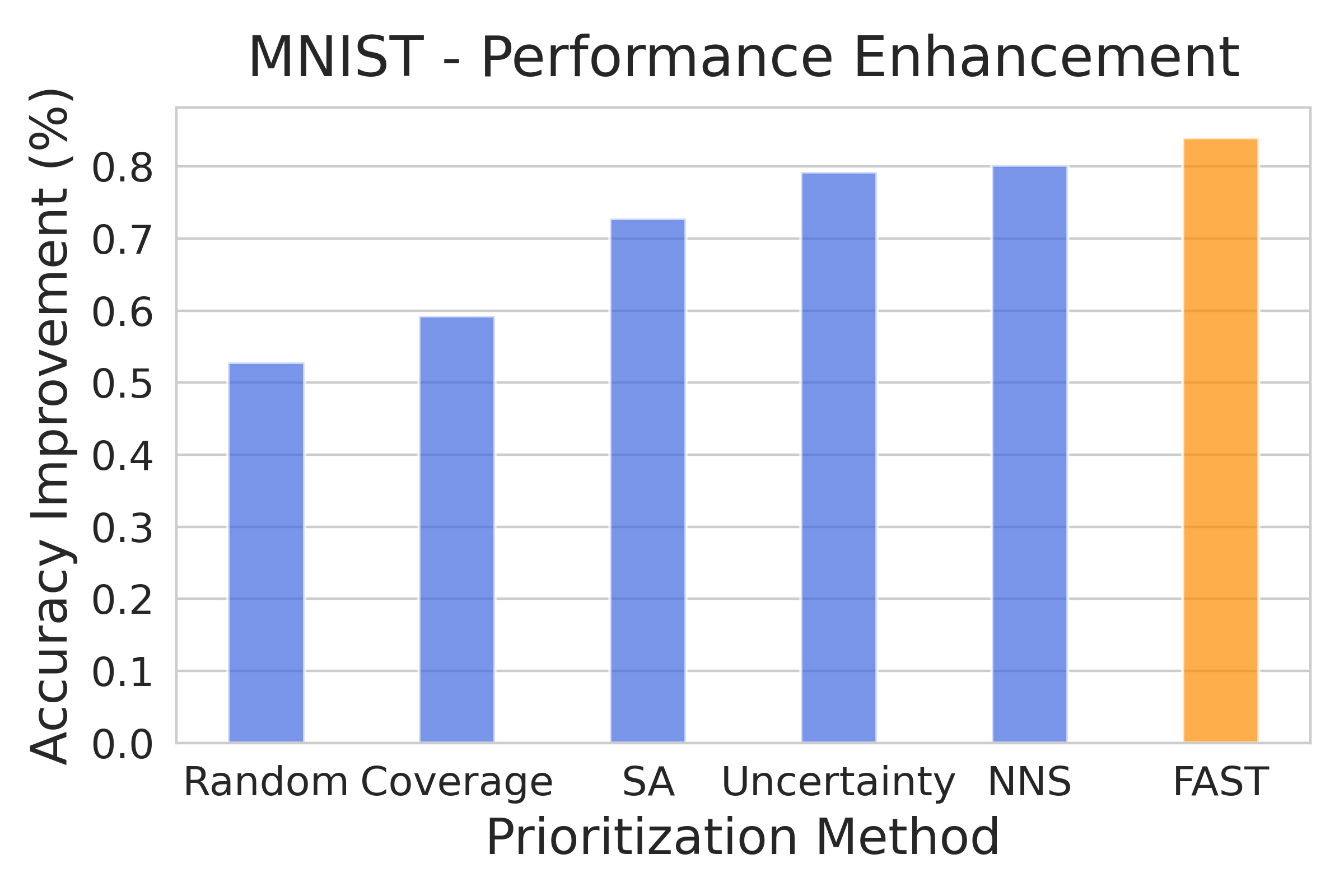}
\hspace{1mm}
\includegraphics[width=0.23\textwidth]{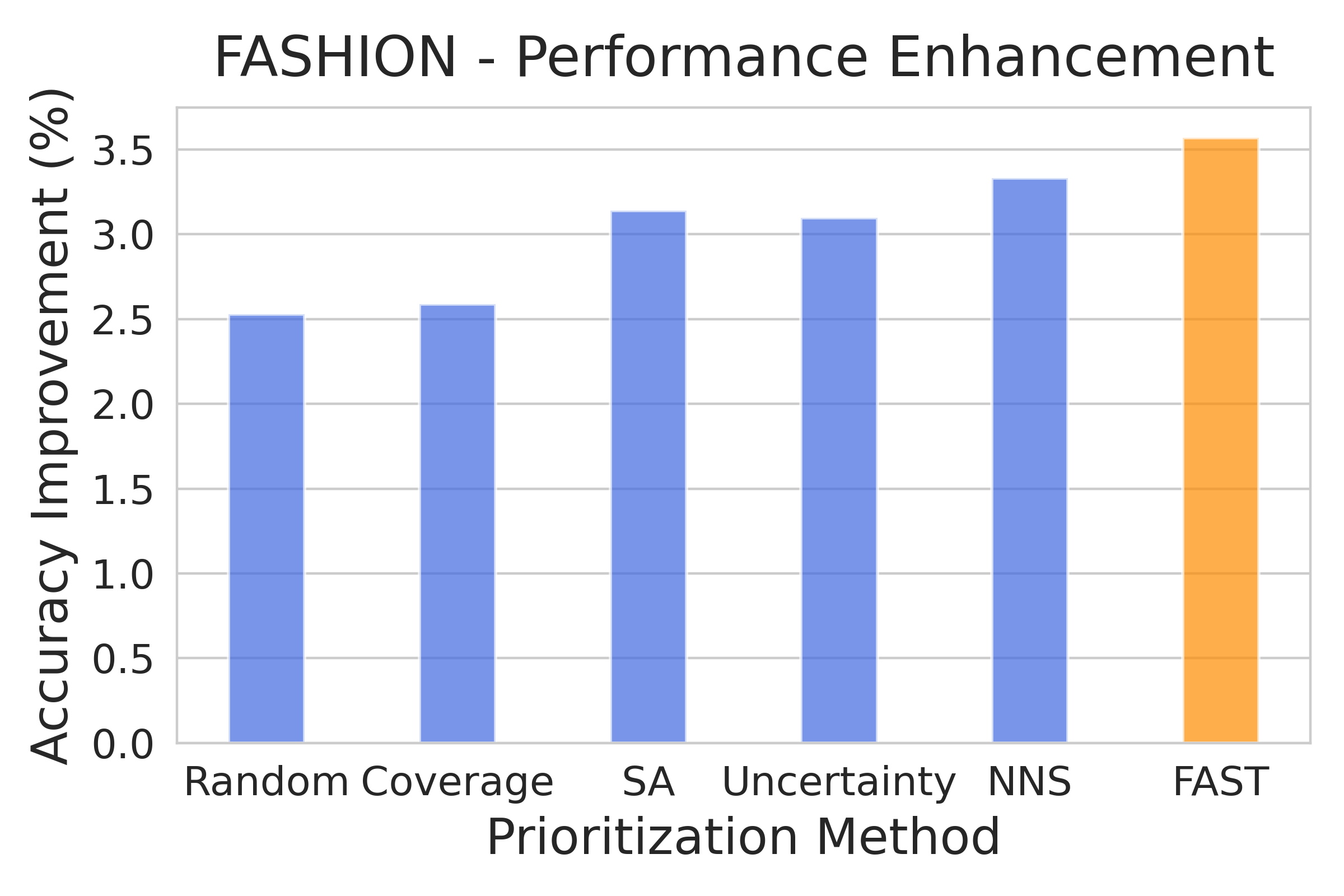}
\hspace{1mm}
\includegraphics[width=0.23\textwidth]{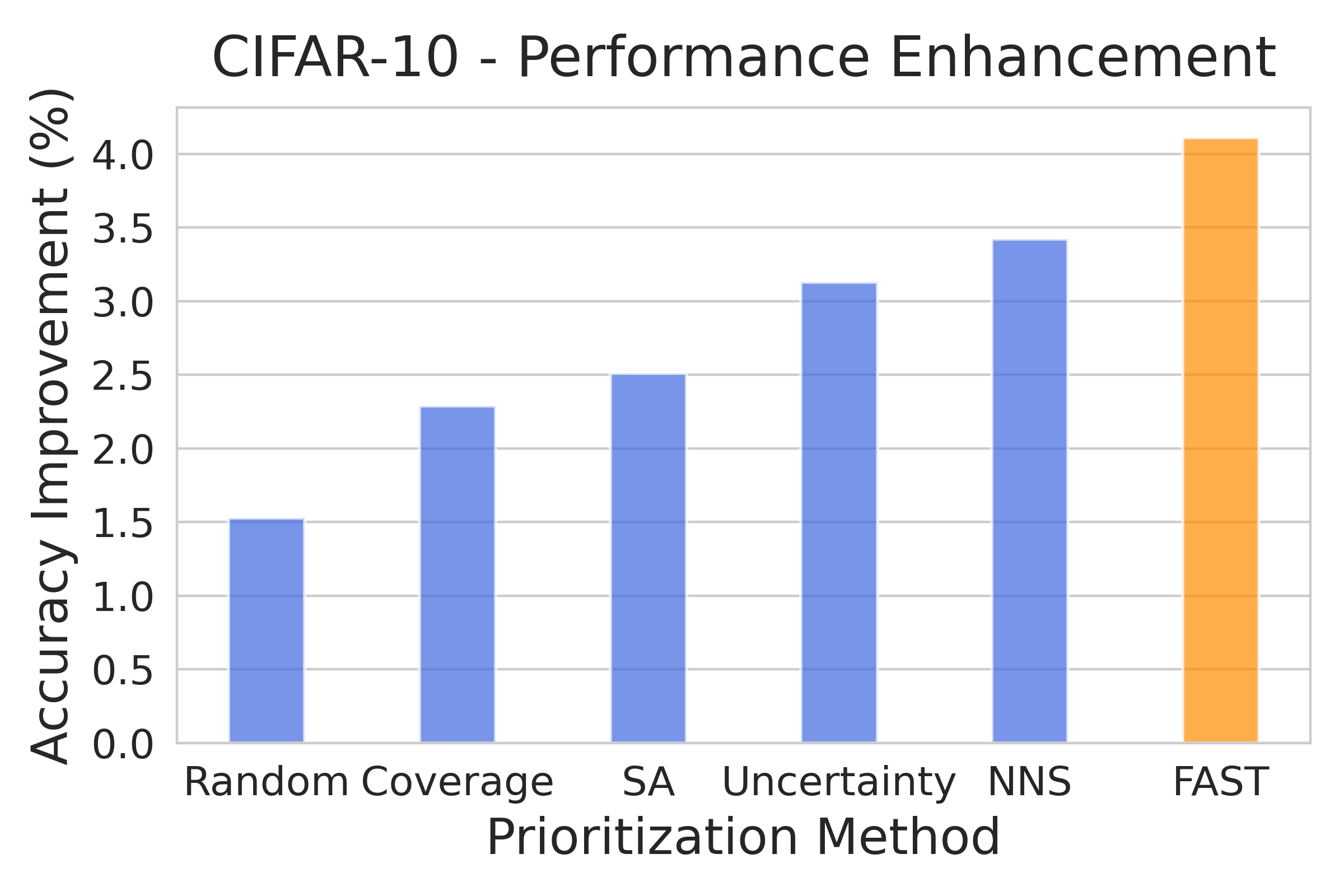}
\hspace{1mm}
\includegraphics[width=0.23\textwidth]{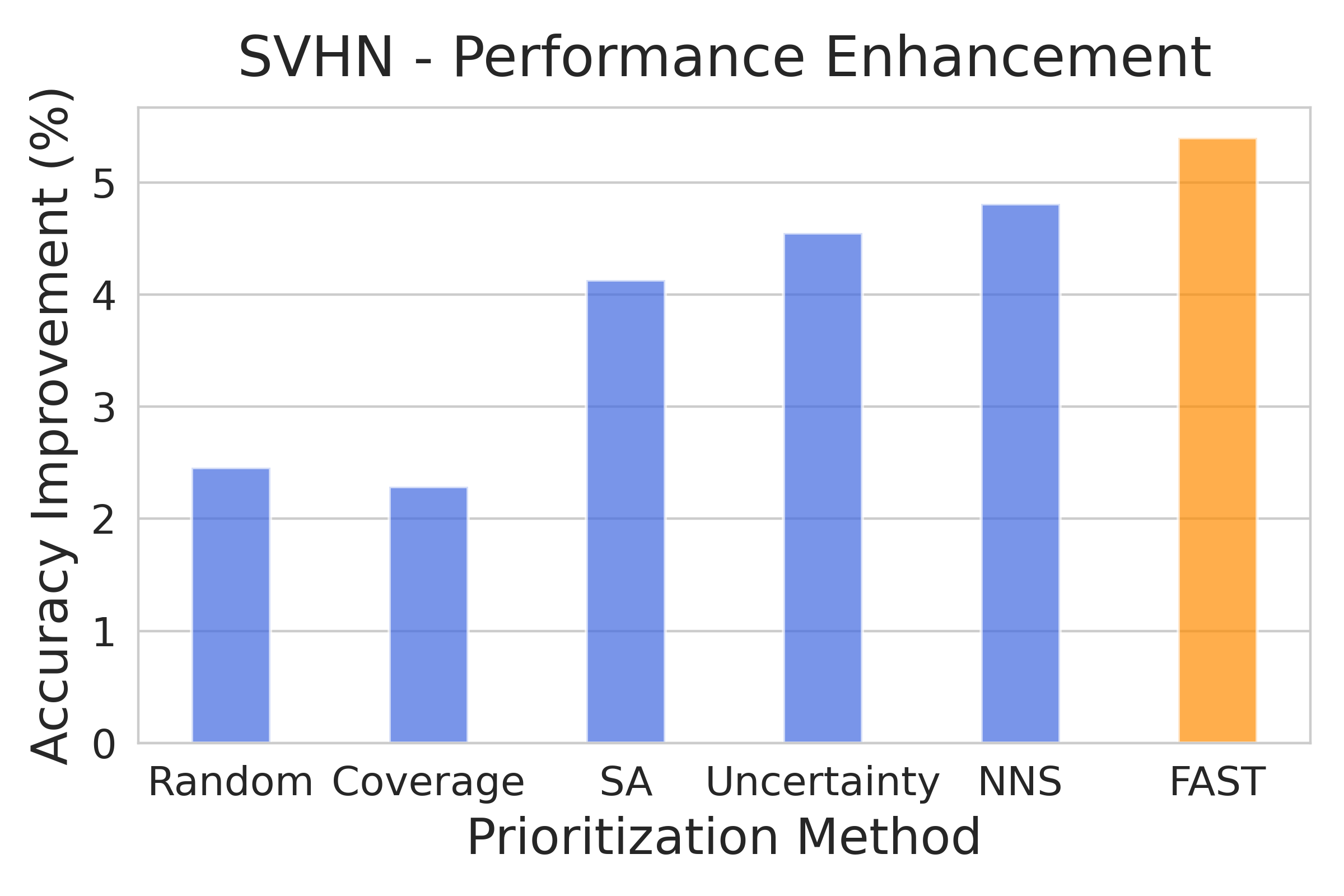}
\vspace*{-2mm}
\caption{The model enhancements guided by test prioritization methods through retraining (with 5\% selection budget).}
\label{fig:enhancement}
\end{figure*}

\subsubsection{RQ4: Is \tool scalable to high-dimensional input and other data domains?}\hspace*{\fill}
\label{rq4}

To investigate the scalability of \tool, we conduct validation experiments on three additional datasets, the Imagenette dataset, which comprises high-dimensional images of $224\times224\times3$ pixels; the GSCmd dataset~\cite{warden2018speech}, which includes audio commands such as `left' and `right'; and the Imdb dataset~\cite{maas2011learning} containing textual reviews for sentiment classification. We perform necessary pre-processing for the audio and text data, and use the RNN model structures (LSTM~\cite{yu2019review} and fully-connected layers) for model training. The results are presented in  Table~\ref{tab:scala}.

Compared to the results on image data, the improvements of NNS and \tool over uncertainty-based methods in the text and audio domains are relatively subtle. We hypothesize that this is due to the more abstract and entangled nature of features extracted in these domains, which affects both NNS and \tool to a certain extent. Nevertheless, \tool still demonstrates a non-trivial improvement over the uncertainty-based methods. This further validates the scalability of \tool to handle complex data and model structures.

\vspace{-1mm}
\begin{tcolorbox}[fonttitle = \bfseries, boxsep=3pt, left=6pt, right=6pt, top=4pt,bottom=4pt]
  \textbf{Answer to RQ4:} \tool is scalable to high-dimensional input and promising for other data domains. Compared to other baselines, it still improves APFD values more effectively.
\end{tcolorbox}

\subsubsection{RQ5: Is \tool able to guide the retraining for model enhancement?}\hspace*{\fill}
\label{rq5}

Prioritizing the test suite and selecting the valuable part to enhance the DNN model performance (e.g., accuracy) by retraining is an important application scenario in DNN testing. To evaluate the effectiveness of \tool as well as other prioritization methods, we select a set of test cases with the same selection budget to retrain the models. For each model, we use the same training settings.
 
Fig.~\ref{fig:enhancement} shows the performance enhancement results of different prioritization methods. We can observe that \tool consistently obtains the highest accuracy improvements across all scenarios. Compared to the random baseline, most other prioritization methods also demonstrate certain level of effectiveness, except for the coverage-based methods. For all the retrained models, \tool achieves 3.47\% accuracy improvements on average, which is 13.36\% higher than NNS. This indicates that \tool can expose more hidden errors and rectify them than the baseline methods.

\begin{tcolorbox}[fonttitle = \bfseries, boxsep=3pt, left=6pt, right=6pt, top=4pt,bottom=4pt]
  \textbf{Answer to RQ5:} \tool can guide the retraining procedure to help improve the model performance. 
\end{tcolorbox}

\subsection{Discussion} 
In this work, we proposed \tool, a lightweight test prioritization designed to enhance the efficiency of DNN testing. The key idea behind \tool is to selectively drop certain noisy features to obtain an alternative probability vector for uncertainty measurement. As a generic method, \tool can be easily applied to various types of DNN structures and support different uncertainty-based metrics, not limited to DeepGini. We also examined the sensitivity of \tool to different hyperparameters through necessary experiments. Some hyperparameters, such as the threshold used for class-wise analysis, are currently set based on empirical observations to identify features that are important or irrelevant for correctly predicting each class. 
We plan to investigate automatic parameter tuning to optimize performance. Additionally, the overconfidence issue explored in this work can be mitigated to some extent by existing model calibration methods~\cite{wang2021rethinking,meronen2024fixing,thulasidasan2019mixup} that involve modifying the model structure or training process. We believe that \tool, like other test prioritization methods, is complementary to model calibration techniques. \tool functions as a plug-and-play module, requiring no modifications to the models themselves.

\subsection{Threats to Validity} 
 
The external threat to validity primarily arises from the datasets, model structures, and baselines used in our study. Following recent replication studies \cite{weiss2022simple,nns}, we considered commonly used public datasets in the literature to ensure a broad evaluation scope. Additionally, we validated the scalability of \tool to high-dimensional image data and other data domains including audio and text. To mitigate threats from DNN models, we employed various well-known architectures with different capacities to limit the impact of model dependency to some extent. Since our method is generic and independent of datasets and model structures, it can be adapted with minor adjustments for further applications. Regarding the prioritization baselines, while there are other works with different approaches~\cite{prima,certpri}, we selected multiple representative methods from different categories and demonstrated the effectiveness and efficiency of \tool.
The internal threat could arise from the implementation of \tool and the baselines used for comparison. To reduce this threat, we utilized available implementations from the authors' released code or re-implemented them according to the original papers. We carefully checked the correctness of our code.
The construct threat lies in the hyper-parameter settings of the experiments. For running the baselines, we followed the standard settings in existing works. Additionally, we conducted necessary ablation studies to provide guidance on setting hyper-parameters for optimal performance of \tool.

\vspace{-0.5mm}
\section{Conclusion}
\label{sec:conclusion}

In this paper, we propose a novel and effective DNN test prioritization method \tool to improve testing efficiency and reduce labeling costs. The key idea behind \tool is to prune noisy information from the irrelevant features which could contribute to the high-confidence errors, which sharpens the uncertainty estimation and improves differentiation between high-confidence correct and erroneous predictions. Extensive experiments on multiple benchmark datasets with various model structures validate the effectiveness, efficiency, and scalability of \tool. Our results demonstrate that \tool consistently enhances the performance of existing uncertainty-based test case selection methods and outperforms state-of-the-art work. As a plug-and-play technique, \tool can be easily integrated with different uncertainty measures.

\vspace{-0.9mm}
\section*{Acknowledgements}
This work was supported by the Key R\&D Program of Zhejiang (Grant No. 2022C01018), the National Science Foundation of China Program (Grant No. 62102359, 62293511, 62088101) and the Program of China Scholarship Council (Grant No. 202306320425). 
MK and XZ received partial support from ELSA: European Lighthouse on Secure and Safe AI project (Grant No. 101070617 under UK guarantee) and the ERC under the European Union’s
Horizon 2020 research and innovation program (FUN2MODEL, Grant No. 834115).

\clearpage
\balance

\bibliographystyle{ACM-Reference-Format}
\bibliography{ref.bib}

\end{document}